%% file: main.tex
\documentclass[conference]{IEEEtran}

\usepackage[utf8]{inputenc}
\usepackage{amsmath,amssymb,amsfonts}
\usepackage{graphicx}
\usepackage{textcomp}
\usepackage[dvipsnames]{xcolor}
\usepackage{hyperref}
\usepackage{booktabs}
\usepackage{multirow}
\usepackage{algorithm}
\usepackage{algorithmic}
\usepackage{listings}
\usepackage{microtype}

\hypersetup{colorlinks=true, citecolor=BlueGreen, linkcolor=OrangeRed, urlcolor=Cyan}

\usepackage{caption}
\usepackage{cite}

\makeatletter
\def\footnoterule{\kern-3\p@
  \hrule \@width 1.4in \kern 2.6\p@} 
\makeatother

\def\BibTeX{{\rm B\kern-.05em{\sc i\kern-.025em b}\kern-.08em
T\kern-.1667em\lower.7ex\hbox{E}\kern-.125emX}}

\IEEEoverridecommandlockouts

\begin{document}

\title{Xe-Forge: Multi-Stage LLM-Powered Kernel Optimization for Intel GPU\thanks{Intel, the Intel logo, Intel Arc, and Intel Xe are trademarks of Intel Corporation or its subsidiaries. Other names and brands may be claimed as the property of others.}}

\author{
  \IEEEauthorblockN{
    Marcin Spoczynski\thanks{Corresponding authors:\newline \href{mailto:marcin.spoczynski@intel.com}{marcin.spoczynski@intel.com}, \href{mailto:daniel.fleischer@intel.com}{daniel.fleischer@intel.com}},
    Daniel Fleischer,
    Moshe Berchansky,
    Gabriela Ben-Melech Stan
  }
  \IEEEauthorblockN{
    Shira Guskin,
    Weilin Xu,
    Adam Siemieniuk,
    Alexander Heinecke
  }
  \IEEEauthorblockN{
    \\
    Intel Corporation
  }
}
\maketitle

\begin{abstract}
  Porting deep learning algorithms to new hardware accelerators requires developers to repeatedly apply the same low-level optimizations — quantization, memory access coalescing, tile size tuning, and architecture-specific workarounds — to every Triton kernel in their code-base. This manual, repetitive effort is a major bottleneck: each kernel demands the same cycle of trial-and-error profiling against hardware constraints that vary across devices, yet the underlying optimization patterns remain largely consistent. We present Xe-Forge, a multi-stage LLM-powered pipeline that automates this process for Intel GPU. Given a functionally correct Triton kernel, the system applies up to nine optimization stages — from algorithmic restructuring and operator fusion through block pointer modernization, GPU-specific tuning, and open-ended discovery — each driven by a Chain-of-Verification-and-Refinement (CoVeR) agent that generates candidates, validates them on real hardware, and iterates on failures. A curated knowledge base encodes Intel GPU constraints (power-of-two warp counts, GRF modes, SLM sizing) that are absent from LLM training data, keeping the model within architecturally valid bounds. We evaluate Xe-Forge on 97 Level-2 KernelBench kernels and Flash Attention on the Intel Arc Pro B70, achieving a $1.17{\times}$ geometric mean speedup over PyTorch eager with 67\% of kernels improving, nine kernels exceeding $5{\times}$ (up to $82{\times}$), and $2$--$13.3{\times}$ speedups on Flash Attention across all tested configurations without regression --- demonstrating that structured domain knowledge with hardware-in-the-loop verification can systematically eliminate the repetitive porting effort that currently gates algorithm deployment on new accelerators.
\end{abstract}

\begin{IEEEkeywords}
  GPU kernel optimization, Triton, Intel GPU, large language models, code generation
\end{IEEEkeywords}

\section{Introduction}
\input{introduction}


\section{Background}
\input{background}
\label{sec:background}
\section{Related Work}
\label{sec:related}

\subsection{LLM-Based Kernel Generation}
\label{sec:related-generation}

LLM-based kernel generation translates a high-level
specification---typically a PyTorch operator or natural-language
description---into a new Triton or CUDA kernel.  We focus on the
technical mechanisms underlying the principal systems.

\paragraph{Closed-loop generation systems.}
The core architectural pattern shared by current generation systems is a
\emph{generate--compile--verify} loop in which the LLM proposes code, a
tool-chain validates it, and error diagnostics are fed back for
refinement.  Systems differ primarily in how they structure this loop and
what verification signals they exploit.  NVIDIA's DeepSeek-R1
integration~\cite{nvidia_deepseekr1_kernels} relies on inference-time
compute scaling: the model generates multiple candidate kernels per
problem, each validated against reference outputs, with chain-of-thought
traces guiding self-correction.  The key technical insight is that
scaling test-time compute (more candidates, longer reasoning chains)
substitutes for additional training.

KernelFalcon~\cite{wang2025kernelfalcon} introduces a \emph{graph
decomposition} front-end: the input PyTorch program is traced into a
data-flow graph, partitioned into fusible sub-graphs via a greedy
clustering heuristic, and each sub-graph is assigned to a dedicated
generation agent.  Agents operate in parallel with independent
verify-and-retry budgets, and a composition layer stitches the resulting
kernels back into a single callable module.  This decomposition enables
the system to handle Level-3 (full-architecture) KernelBench problems
that monolithic generation cannot address.

KernelEvolve~\cite{liao2025kernelevolve} replaces single-shot generation
with \emph{graph-based evolutionary search}: a population of kernel
variants is maintained, and at each generation step the LLM produces
mutations conditioned on retrieval-augmented context drawn from a
curated corpus of high-performance Triton patterns.  Fitness is
determined by wall-clock execution time on the target device.  The
system maintains separate populations per hardware backend (NVIDIA, AMD,
MTIA), allowing hardware-specific evolution without cross-contamination
of device-specific idioms.

Astra~\cite{wei2025astra} coordinates multiple specialized agents
(generator, tester, profiler, planner) that collaborate through
iterative refinement, each contributing a distinct verification signal.
STARK~\cite{dong2025stark} reframes kernel optimization as strategic
search over a persistent tree memory, where nodes store candidates with
their runtime, correctness, and compiler diagnostics, enabling
backtracking and branch-and-bound pruning.
KernelSkill~\cite{sun2026kernelskill} introduces a dual-level memory
architecture---long-term reusable expert skills paired with short-term
backtracking prevention---achieving $5.4{\times}$/$2.8{\times}$/$1.9{\times}$
speedups on KernelBench~L1/L2/L3.
AutoKernel~\cite{jaber2026autokernel} takes a whole-model view: it
profiles a PyTorch model to identify bottleneck kernels via Amdahl's
law, then iteratively optimizes Triton or CUDA~C++ kernels through
hundreds of edit--benchmark--keep/revert cycles.

VibeTensor~\cite{xu2026vibetensor} extends LLM-driven code synthesis
beyond individual kernels to an entire deep learning runtime.  The
system uses a hierarchical agent architecture: a planning agent
decomposes the runtime into components (operator library, memory
allocator, scheduler), specialist agents generate each component, and
an integration agent resolves cross-component interfaces.  Correctness
is enforced through a test harness that validates each component
independently before system-level integration.

\paragraph{Benchmarks.}
KernelBench~\cite{ouyang2025kernelbench} organizes evaluation into three
difficulty levels---single operators, operator compositions, and full
model architectures---with each problem specifying a reference PyTorch
implementation, input generators, and correctness tolerances
(\texttt{torch.allclose} with configurable \texttt{rtol}/\texttt{atol}).
A kernel passes only if it produces correct outputs \emph{and} compiles
and runs without error; performance is measured but not gated.
TritonBench~\cite{li2025tritonbench} focuses specifically on Triton
operator generation and introduces finer-grained metrics including
syntactic validity rate, compilation success rate, and performance
relative to \texttt{torch.compile} baselines, enabling diagnosis of
where in the generation pipeline failures occur.

\paragraph{Specialized training and RL.}
KernelLLM~\cite{fisches2025kernelllm} fine-tunes a base model on
curated (PyTorch, Triton) pairs extracted from open-source repositories,
using a curriculum that progresses from simple elementwise operators to
multi-stage reductions.  AutoTriton~\cite{li2025autotriton} formulates
kernel generation as a reinforcement learning problem: the policy
network generates Triton code token-by-token, and the reward combines
a binary compilation signal with a continuous performance score
normalized against a reference implementation.
Kevin~\cite{baronio2025kevin} extends RL to multi-turn CUDA generation,
modeling the generate--compile--profile cycle as a Markov decision
process where each turn's observation includes compiler diagnostics and
profiler output from the previous attempt, allowing the policy to learn
error-recovery strategies.  CUDA Agent~\cite{dai2026cudaagent} scales
this paradigm with large-scale agentic RL, synthetic data generation,
and a skill-augmented CUDA development environment, achieving
100\%/100\%/92\% faster-than-\texttt{torch.compile} rates on
KernelBench L1/L2/L3.

\paragraph{Positioning.}
These systems share a common formulation: the input is a functional
specification and the output is a \emph{new} kernel.  With the
exceptions of KernelEvolve (AMD, MTIA) and GEAK (AMD), they target
NVIDIA hardware.  None addresses Intel GPU, and none takes an existing
kernel as input for architecture-specific refinement.

\subsection{LLM-Based Kernel Optimization}
\label{sec:related-optimization}

A complementary line of work takes an existing, correct kernel as input
and improves it for a target architecture.

\paragraph{GEAK and GEAK-v2 (AMD Instinct).}
AMD's GEAK framework~\cite{wang2025geak} targets Triton kernel
optimization for AMD Instinct GPUs using a Reflexion-style
architecture: after each failed generation attempt, the agent produces
a structured self-critique prepended to the next prompt, creating an
explicit error-correction trajectory.  GEAK-v2~\cite{geakv2_2025}
augments this with evolutionary search over kernel populations and
integrates \texttt{rocprof} profiler traces as feedback signals,
enabling the agent to reason about hardware counter data (occupancy,
memory bandwidth utilization) rather than execution time alone,
achieving up to $2.59\times$ speedup on AMD hardware.

\paragraph{Apex (AMD ROCm).}
AMD's Apex~\cite{apex2025} is an RL environment that tasks an LLM
agent---Claude Code or OpenAI Codex---with optimizing GPU kernels
drawn from a registry of 19 open-source LLMs and 12 kernel types
(Flash Attention prefill, paged attention decode, fused MoE, RMS
norm, RoPE embedding, and others) for AMD Instinct hardware (default:
MI355X / gfx950).  The agent operates in a sandbox populated with
ROCm source repositories, architecture documentation, and
best-practice guides, and is scored via the AgentKernelArena
formula: compilation (+20~pts), correctness (+100~pts), and speedup
($\times$100~pts).  Grading is delegated to Magpie, AMD's kernel
benchmarking framework.  Apex also exposes a model-level scoring mode
that grounds optimization decisions in end-to-end inference
throughput, rather than isolated kernel performance.

\paragraph{AVO (NVIDIA Blackwell).}
NVIDIA's Agentic Variation Operators (AVO)~\cite{avo2026} replace the
fixed mutation and crossover steps of classical evolutionary search
with a self-directed coding agent that subsumes sampling, generation,
and evaluation into a single autonomous loop.  Applied to multi-head
attention on the Blackwell B200 GPU over 7 days of continuous
autonomous evolution, AVO discovers kernels that surpass cuDNN by up
to 3.5\% and FlashAttention-4 by up to 10.5\%, reaching 1668~TFLOPS
at BF16 precision.  The agent operates on CUDA and PTX source
directly, consulting hardware documentation and profiler output
iteratively; the discovered optimizations---branchless accumulator
rescaling, correction/MMA pipeline overlap, register rebalancing
across warp groups---reflect genuine micro-architectural reasoning
rather than parameter sweeping.  Critically, optimizations discovered
on MHA transfer to grouped-query attention in only 30 minutes of
additional autonomous adaptation, demonstrating cross-variant
generalization.

\paragraph{SOL-ExecBench.}
SOL-ExecBench~\cite{solexecbench2026} addresses a measurement gap in
the optimization literature: existing benchmarks evaluate kernels
against software baselines (e.g.\ PyTorch eager) rather than
hardware-grounded performance limits.  SOL-ExecBench introduces a
speed-of-light (SOL) metric that measures kernel efficiency as a
fraction of the theoretical hardware maximum, covering frontier and
emerging architectures, FP8 precision formats, and both training and
inference workloads.  Its evaluation infrastructure is designed to be
robust to adversarial optimization---a failure mode also relevant to
our setting (Section~\ref{sec:discussion})---and it explicitly
distinguishes between kernels that are fast relative to software
baselines and kernels that are fast relative to hardware limits.

\paragraph{TritonForge (NVIDIA).}
TritonForge~\cite{li2025tritonforge} is a profiling-guided framework
that feeds NVIDIA Nsight Compute metrics directly to an LLM agent,
which iteratively rewrites Triton kernels to address the specific
bottlenecks identified by the profiler.  By grounding optimization
decisions in hardware counter data rather than wall-clock time alone,
TritonForge achieves targeted improvements on memory-bound kernels.

\paragraph{KernelFoundry (SYCL).}
KernelFoundry~\cite{wiedemann2026kernelfoundry} applies MAP-Elites
quality-diversity search with meta-prompt evolution and template-based
parameter optimization, reporting $2.3{\times}$ average speedup on
KernelBench.  Notably, it targets SYCL as its backend, making it the
closest prior work to our Intel GPU setting, though it does not
incorporate Intel-specific hardware constraints or a curated knowledge
base.

\paragraph{Positioning.}
These systems collectively demonstrate that LLM-driven kernel
optimization can match or exceed expert-engineered implementations
across NVIDIA and AMD hardware.  GEAK and Apex target AMD ROCm; AVO
targets NVIDIA Blackwell; KernelFoundry targets SYCL; none targets
Intel Xe GPU with architecture-specific constraints.  AVO and Apex
use open-ended single-agent loops; GEAK and KernelFoundry use
evolutionary search; our pipeline decomposes optimization into nine
dependency-ordered stages each with dedicated verification criteria,
trading flexibility for reproducibility and predictable cost.  SOL-ExecBench's hardware-grounded evaluation
methodology motivates our use of AI~Bench's hardware-event timing
rather than software-baseline speedup as the primary performance
signal.

\subsection{Hardware-Specific Optimization}
\label{sec:related-hardware}

Before LLM-based approaches, hardware-specific kernel optimization was
dominated by autotuning frameworks that search over parameterized
configuration spaces using classical optimization techniques.

\paragraph{Traditional autotuners.}
OpenTuner~\cite{ansel2014opentuner} introduced an extensible framework
for domain-specific autotuning that ensembles disparate search
techniques---differential evolution, simulated annealing, Nelder-Mead,
and others---allocating a larger testing budget dynamically to
techniques that perform well. Rather than committing to a single search
strategy, OpenTuner treats the choice of optimizer as itself an
optimization problem, demonstrating up to $2.8\times$ speedup over
prior techniques across 16 benchmarks.  Kernel Tuner~\cite{vanwerkhoven2019kerneltuner}
applies a similar philosophy specifically to GPU kernels, supporting
CUDA, OpenCL, and HIP backends with a Python interface and a suite of
global optimization solvers.  It demonstrates that search-optimizing
strategies---particularly basin hopping combined with gradient-free
solvers---can find near-optimal GEMM configurations significantly faster
than exhaustive enumeration.

Ansor~\cite{zheng2020ansor}, integrated into Apache TVM as
\texttt{tvm.auto\_scheduler}, moves beyond parameter search to
\emph{program generation}: rather than tuning a fixed kernel
implementation, Ansor samples tensor programs from a hierarchical
representation of the full optimization search space (tiling,
parallelism, unrolling, and memory layout decisions), then fine-tunes
sampled programs with evolutionary search guided by a learned cost
model.  This template-free approach avoids the manual schedule
engineering required by AutoTVM and improves execution performance
by up to $3.8\times$, $2.6\times$, and $1.7\times$ over prior
state-of-the-art on Intel CPU, ARM CPU, and NVIDIA GPU respectively.

\paragraph{Limitations of traditional autotuning for Intel GPU.}
These frameworks share a common assumption: the optimization search
space is defined by numerical parameters (tile sizes, block dimensions,
unroll factors) within a fixed kernel structure.  They cannot alter
the kernel's algorithmic structure, change its data types, fuse
multiple kernels, or apply architecture-specific code transformations
such as block pointer modernization or GRF mode annotation.
Furthermore, Ansor and Kernel Tuner target NVIDIA and AMD hardware
with well-established cost models; extending them to Intel GPU requires
authoring new hardware back-ends, cost models, and schedule rules---an
engineering effort comparable to writing the optimized kernels by hand.
Intel's oneAPI toolkit and SYCL programming guides provide
architecture-specific tuning guidance for Xe GPU cores (EU counts,
subslice memory hierarchy, GRF modes, SLM capacity), but these are
documentation resources rather than automated optimization systems:
they inform the human engineer but do not automate the tuning loop.

\paragraph{Gap.}
No existing autotuning system combines LLM-driven code transformation
with hardware-in-the-loop verification targeting Intel GPU.
Traditional autotuners optimize parameters within a fixed kernel;
LLM-based generation systems target NVIDIA or AMD hardware and produce
new kernels rather than refining existing ones.  Xe-Forge fills
this gap by augmenting a classical autotuning stage
(Section~\ref{sec:stages}) with eight upstream LLM-driven
transformation stages that restructure the kernel before any parameter
search begins, encoding Intel GPU constraints in a curated knowledge
base that substitutes for the hardware-specific cost models that
traditional autotuners require.

\section{System Design}
\label{sec:system}

\subsection{Pipeline Architecture}
\label{sec:pipeline-arch}

Xe-Forge is structured as a multi-stage pipeline
that takes a functionally correct Triton kernel as input and produces
an optimized variant targeting Intel GPU.  The pipeline comprises
an analysis stage followed by up to nine optimization stages whose
execution order is determined by an LLM-based planner subject to
hard dependency constraints:

\begin{enumerate}
  \item \textbf{Analysis} --- detect optimization opportunities and
    classify issues by type,
  \item \textbf{Algorithmic optimization} --- reduce FLOPs and memory
    accesses by exploiting mathematical structure: matrix symmetry,
    common sub-expression elimination, loop-invariant hoisting,
    algebraic simplification of fused computations, and tree
    reductions in place of serial accumulation,
  \item \textbf{Discovery} --- apply novel, open-ended optimizations
    not covered by any named stage (e.g.\ algebraic elimination of
    an entire GEMM, caching of weight statistics across forward
    passes),
  \item \textbf{Dtype fix} --- eliminate unnecessary precision
    (e.g.\ \texttt{float64} $\to$ \texttt{float32}, insert
    mixed-precision casts),
  \item \textbf{Fusion} --- merge separate kernels or elementwise
    chains into a single launch,
  \item \textbf{Memory access} --- improve coalescing, prefetching,
    and data layout,
  \item \textbf{Block pointers} --- replace manual pointer arithmetic
    with the \texttt{tl.make\_block\_ptr} API,
  \item \textbf{Persistent kernel} --- restructure loops for
    cross-iteration data reuse,
  \item \textbf{GPU-specific} --- apply Intel-specific tuning (warp
    count, GRF mode, tile sizing, GROUP\_SIZE\_M swizzling),
  \item \textbf{Autotuning} --- generate a
    \texttt{@triton.autotune} configuration grid derived from
    hardware parameters and problem shape.
\end{enumerate}

\paragraph{Stage ordering and planner.}
Rather than executing stages in a fixed order, the pipeline employs an
LLM-based \emph{planner} that determines the optimal stage sequence for
each kernel based on the detected issues and their interactions.  The
planner operates under a set of hard dependency constraints that encode
the principle of \emph{decreasing semantic scope}:

\begin{itemize}
  \item \textbf{Algorithmic/Discovery $\to$ Dtype, Fusion.}
    Structural rewrites (algorithmic simplification, open-ended
    discovery) run first because they may eliminate entire operator
    chains, rendering downstream dtype or fusion changes unnecessary.
  \item \textbf{Dtype $\to$ Fusion.}
    Fusing two \texttt{float64} kernels then converting produces
    suboptimal mixed-precision code; fixing dtypes first avoids
    rework.
  \item \textbf{Memory access $\to$ Block pointers.}
    Coalescing and layout optimization must stabilize before the
    addressing mechanism is rewritten with \texttt{tl.make\_block\_ptr}.
  \item \textbf{Fusion, Block pointers $\to$ GPU-specific.}
    Intel-specific tuning (tile sizes, warp counts, GRF mode)
    depends on the final kernel structure.
  \item \textbf{GPU-specific $\to$ Autotuning.}
    The autotune configuration grid is derived from the
    GPU-specific parameters; tuning before they are set would be
    wasted effort.
\end{itemize}

\noindent Within these constraints, the planner reorders stages to
minimize redundant passes---for example, skipping fusion entirely if
algorithmic optimization already collapsed the operator chain.  If the
planner cannot determine an order (e.g.\ LLM failure), it falls back
to the default sequence: Algorithmic $\to$ Discovery $\to$ Dtype
$\to$ Fusion $\to$ Memory $\to$ Block Pointers $\to$ Persistent
Kernel $\to$ GPU-Specific $\to$ Autotuning.

\paragraph{Issue-driven skip logic.}
Not every kernel requires all nine optimization stages.  The analysis
stage (Section~\ref{sec:stages}) invokes an LLM-based analyzer that
classifies detected issues by type (e.g.\ \texttt{dtype\_float64},
\texttt{unfused\_kernels}, \texttt{manual\_pointer\_arithmetic},
\texttt{open\_ended});
each issue type maps deterministically to exactly one downstream
stage via a fixed routing table (with dynamic registration for
custom issue types).  The planner collects the set of stages
that have at least one associated issue and produces an ordered
subset respecting the hard dependency constraints: if the analyzer
detects no fusion opportunities, the fusion stage is skipped
entirely.  This selective execution reduces both LLM inference cost
and the risk of unnecessary code churn in stages where no
improvement is expected.

\paragraph{Re-analysis between stages.}
After each optimization stage completes, the pipeline re-invokes the
analyzer on the current kernel code.  This serves two purposes.
First, a successful optimization may resolve issues that span multiple
stages---for example, a fusion pass that also eliminates a redundant
dtype cast---allowing subsequent stages to be skipped.  Second, an
optimization may introduce new issues (e.g.\ a fusion pass that
increases register pressure), which the downstream stages can then
address.  The re-analysis step keeps the issue list synchronized with
the actual kernel state rather than relying on the initial analysis
throughout.

\paragraph{Best-of-$k$ selection.}
To mitigate the variance inherent in LLM-based code generation, the
pipeline supports running the full stage sequence $k$ times
independently, producing $k$ candidate optimized kernels.  Each
candidate is benchmarked on the target device, and the one with the
highest measured speedup is selected as the final output.  In
practice, $k{=}1$ is sufficient for most kernels; we use $k{>}1$ only
when targeting maximum performance on high-value kernels where
additional LLM inference cost is justified.

\begin{figure}[t]
  \centering
  \includegraphics[width=\columnwidth]{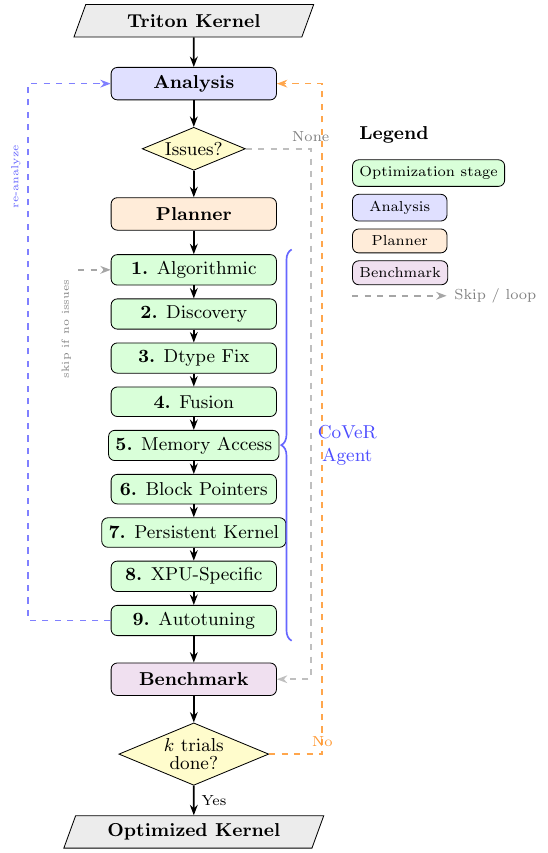}
  \caption{Pipeline architecture.  The analysis stage classifies
    issues; the planner determines an execution order for the nine
    optimization stages subject to hard dependency constraints.
    Stages with no associated issues are skipped.  After each stage,
    the kernel is re-analyzed to update the issue set.  The CoVeR
    agent (Section~\ref{sec:cover}) executes each optimization stage
  internally.}
  \label{fig:pipeline_overview}
\end{figure}

\subsection{Chain-of-Verification-and-Refinement (CoVeR)}
\label{sec:cover}

Each optimization stage in the pipeline is executed by a
\emph{Chain-of-Verification-and-Refinement} (CoVeR) agent---a custom
DSPy module that iteratively generates candidate optimizations,
validates them through a multi-level verification tool, and refines
based on concrete error feedback.  CoVeR is the mechanism by which
the pipeline converts a detected issue into a verified, faster kernel.

\paragraph{Module design.}
CoVeR extends the DSPy \texttt{Module} base class and is parameterized
by three components: (i)~a \emph{signature} that defines the
optimization task's input and output fields (original code, current
  code, stage, issues, knowledge patterns, GPU configuration
$\to$ optimized code), (ii)~a set of \emph{tools} that the agent can
invoke---in our system, a single \texttt{compile\_and\_verify}
tool---and (iii)~a \emph{success sentinel}, a distinguished string
that, when returned by a tool, signals that the current candidate
satisfies all verification checks.  Internally, the module maintains a
\emph{trajectory}: a growing key--value log of the agent's thoughts,
tool invocations, and observations across iterations.  At each step,
the full trajectory is formatted and included in the LLM prompt,
giving the model access to its own prior reasoning and the specific
errors it has encountered.

\paragraph{Generate--verify--refine loop.}
Algorithm~\ref{alg:cover} summarizes the CoVeR execution loop.  At
each iteration $i \in \{0, \ldots, T{-}1\}$, the agent receives the
task inputs together with the accumulated trajectory and produces two
outputs: a \texttt{next\_thought} (free-form reasoning about the
current state and intended changes) and the task output fields
(the optimized code).  The thought is recorded in the trajectory, and
the optimized code is passed to every registered tool.  Each tool
returns an observation---either the success sentinel or a structured
error message---which is appended to the trajectory.  If any tool
returns the success sentinel, the loop terminates immediately and
returns the current outputs.  Otherwise, the error observations
become part of the context for the next iteration, allowing the agent
to diagnose and correct specific failures.  If the loop exhausts
all $T$ iterations without success, a fallback
\texttt{ChainOfThought} extractor is invoked on the final trajectory
to produce a best-effort output.

\begin{algorithm}[t]
  \caption{CoVeR execution loop}
  \label{alg:cover}
  \begin{algorithmic}[1]
    \REQUIRE Task inputs $\mathbf{x}$, tools $\mathcal{T}$, max iterations $T$, success sentinel $s$
    \STATE $\tau \gets \emptyset$ \COMMENT{trajectory}
    \FOR{$i = 0$ \TO $T-1$}
    \STATE $(\textit{thought}_i,\; \mathbf{y}_i) \gets \textsc{LLM}(\mathbf{x},\; \tau)$
    \STATE $\tau[\texttt{thought\_}i] \gets \textit{thought}_i$
    \FOR{each tool $t \in \mathcal{T}$}
    \STATE $o_{i,t} \gets t(\mathbf{y}_i)$
    \IF{$o_{i,t} = s$}
    \RETURN $\mathbf{y}_i$
    \ENDIF
    \ENDFOR
    \STATE $\tau[\texttt{obs\_}i] \gets \{o_{i,t}\}_{t \in \mathcal{T}}$
    \ENDFOR
    \RETURN \textsc{Fallback}$(\mathbf{x},\; \tau)$ \COMMENT{ChainOfThought extraction}
  \end{algorithmic}
\end{algorithm}

\paragraph{Multi-level verification.}
The \texttt{compile\_and\_verify} tool implements a four-level
verification cascade; each level gates entry to the next, and
the first failure produces a diagnostic message that is returned
to the agent:

\begin{enumerate}
  \item \textbf{Syntax.} The candidate is parsed via Python's
    \texttt{ast.parse}.  On failure, the agent receives the line
    number, error type, and the offending source line.

  \item \textbf{Structure.} The tool checks for required components:
    Triton and \texttt{triton.language} imports, a
    \texttt{@triton.jit}-decorated kernel function, and a
    \texttt{Model} wrapper class.  It also validates
    hardware constraints---\texttt{num\_warps} must be a power of
    two, block dimensions must be powers of two and at most~256.
    Violations produce specific remediation instructions
    (e.g.\ ``\texttt{INVALID num\_warps=24: Must be a power
    of 2. Valid values: 1, 2, 4, 8, 16, 32}'').

  \item \textbf{Correctness.} Both the original and optimized kernels
    are instantiated with identical weights (seeded initialization
    followed by explicit \texttt{state\_dict} copy) and executed on
    the same inputs.  Outputs are compared via
    \texttt{torch.allclose} with configurable tolerances
    ($\texttt{rtol}$, $\texttt{atol}$).  On mismatch, the agent
    receives a structured message identifying likely causes
    (wrong strides, transposed loads, missing boundary checks).

  \item \textbf{Performance.} The optimized kernel is benchmarked
    against the original using hardware-event timing with
    warmup, L2 cache flushing, and synchronization barriers.
    If the optimized kernel is slower, the agent receives both
    timings, TFLOPS figures, and suggestions for alternative
    optimization strategies.  Only when the optimized kernel is
    both correct and faster does the tool return the success
    sentinel.
\end{enumerate}

\noindent
This cascade ensures that the agent addresses errors in order of
severity: there is no value in benchmarking a kernel that does not
parse, and no value in measuring performance of a kernel that
produces wrong outputs.

\paragraph{Trajectory management.}
As iterations accumulate, the trajectory can exceed the LLM's
context window.  CoVeR handles this with a truncation policy:
when a context-length error is caught, the four oldest trajectory
entries (one thought, tool name, tool arguments, and observation)
are removed, preserving the most recent error feedback that is
most relevant for the next attempt.  If the trajectory cannot be
truncated further (only one tool call remains), the agent raises
an error rather than operating without diagnostic context.

\paragraph{Fallback behavior.}
When all $T$ iterations are exhausted without the success sentinel,
the pipeline does not discard the work.  A \texttt{ChainOfThought}
extractor receives the full trajectory and produces a best-effort
optimized kernel.  The pipeline then independently verifies this
output through the same four-level cascade.  If verification fails,
the stage returns the \emph{original} code unchanged, ensuring the
pipeline never degrades a working kernel.  Failed kernels are
additionally dumped to disk with stage and timestamp metadata for
post-hoc debugging.


\subsection{Optimization Stages}
\label{sec:stages}

Each optimization stage addresses a distinct class of performance
issues, identified by the analysis stage and routed via the
issue-to-stage mapping described in Section~\ref{sec:pipeline-arch}.
Below we detail what each stage detects, transforms, and verifies.

\paragraph{(1) Analysis.}
The analysis stage does not modify the kernel; it produces a
structured inventory of optimization opportunities that drives all
subsequent stages.  An LLM-based analyzer receives the Triton kernel
source, an optional PyTorch reference implementation, the full set of
optimization patterns from the knowledge base, and problem context
(input shapes, FLOP count, target dtype).  It returns a list of
typed issues, each annotated with a severity score (1--5), a textual
description, a suggested fix, and an estimated speedup range.
Issue types are drawn from a fixed taxonomy of 30+ categories
(Table~\ref{tab:issue_types}) spanning all downstream stages.
The severity scores are advisory---they inform the LLM's
prioritization within a stage but do not gate stage execution.

\begin{table}[t]
  \centering
  \small
  \begin{tabular}{@{}p{0.22\columnwidth}p{0.70\columnwidth}@{}}
    \toprule
    \textbf{Stage} & \textbf{Example} \\
    \midrule
    Dtype fix    & \texttt{float64} accumulator, redundant cast \\
    Fusion       & Separate launch per op, chainable pointwise ops \\
    Memory       & Strided column loads, no bounds on \texttt{tl.load} \\
    Block ptr    & Hand-computed offsets, boolean boundary flag \\
    Persistent   & Relaunchable reduction \\
    GPU          & \texttt{num\_warps}${\ne}32$, no GRF annotation, missing \texttt{GROUP\_SIZE\_M} \\
    \bottomrule
  \end{tabular}
  \caption{Issue taxonomy (representative subset). Each type maps
  deterministically to one optimization stage.}
  \label{tab:issue_types}
\end{table}

\paragraph{(2) Algorithmic optimization.}
The algorithmic stage operates at a higher level of abstraction than any
subsequent stage.  Whereas later stages preserve the kernel's
computational structure and tune its implementation parameters, the
algorithmic stage is permitted to restructure the computation
itself---reducing the number of floating-point operations or memory
accesses by exploiting mathematical properties of the problem, while
maintaining numerical equivalence with the original kernel.

\paragraph{(3) Discovery.}
The discovery stage handles \texttt{open\_ended} issues---novel
optimization opportunities that do not fit any named issue category.
When the analyzer identifies such an opportunity, it must provide a
detailed proposal specifying (a)~exactly what changes, (b)~why the
transformation is mathematically valid, (c)~a before/after code sketch,
and (d)~an estimated speedup with reasoning.  Examples include
rewriting \texttt{sum(x @ W.T, dim=1)} as \texttt{x @ W.sum(dim=0)}
to eliminate an $O(MNK)$ GEMM, or caching weight statistics that are
recomputed every forward pass.  The hard dependency constraints ensure
Discovery runs before Dtype and Fusion, since structural rewrites may
eliminate entire operator chains.  Successfully promoted discoveries
are logged for potential integration as named issue types in future
knowledge-base updates.

\paragraph{(4) Dtype fix.}
This stage targets three issue categories: \texttt{float64}
arithmetic that wastes half the hardware's throughput
(\texttt{dtype\_float64}), unnecessary precision in intermediate
computations (\texttt{dtype\_precision}), and redundant input type
conversions (\texttt{dtype\_input\_conversion}).  The CoVeR agent
receives knowledge-base patterns showing how to convert
\texttt{float64} inputs and accumulators to the target precision
(typically \texttt{float16} or \texttt{bfloat16} for
inputs/outputs, \texttt{float32} for accumulation) while inserting
explicit casts at numerically sensitive boundaries.  The
verification tool's correctness level is critical here: a dtype
change that silently degrades output quality is caught by the
\texttt{torch.allclose} comparison with the original kernel's
outputs.

\paragraph{(5) Fusion.}
The fusion stage addresses six issue categories, ranging from
straightforward cases (\texttt{unfused\_kernels}: separate kernel
launches that share data) to nuanced ones
(\texttt{fusion\_register\_pressure}: a fusion that would exceed the
  register file capacity, and \texttt{fusion\_replaces\_vendor}: a
  fused Triton kernel that replaces a vendor library call such as
\texttt{torch.mm}).  The agent is provided with patterns
demonstrating how to merge elementwise chains, fold reductions with
subsequent pointwise operations, and eliminate dead operations
(\texttt{fusion\_noop}).  Because fusion changes the kernel's
computational structure, the correctness verification level is
especially important: the fused kernel must produce bitwise-close
outputs under the same tolerance as the original multi-kernel
sequence.

\paragraph{(6) Memory access.}
This stage optimizes how data moves between global memory and
registers.  The issue taxonomy covers coalescing
(\texttt{uncoalesced\_access}: loads that span non-contiguous
addresses), boundary safety (\texttt{missing\_boundary\_check}:
\texttt{tl.load} without masking, risking out-of-bounds reads),
host--device synchronization in hot paths
(\texttt{device\_host\_sync}: \texttt{.item()} calls that stall
the pipeline), non-contiguous inputs
(\texttt{non\_contiguous\_input}), and register-lifetime issues
(\texttt{long\_liveness}, \texttt{high\_register\_pressure}).
Knowledge-base patterns encode coalescing-friendly access orderings,
prefetch insertion points, and strategies for reducing live register
counts by reordering loads and stores.

\paragraph{(7) Block pointers.}
Triton's \texttt{tl.make\_block\_ptr} API provides a structured
alternative to manual pointer arithmetic, enabling the compiler to
infer alignment and contiguity and emit more efficient memory
instructions.  This stage detects kernels that compute tile
addresses manually (\texttt{manual\_pointer\_arithmetic}) and
rewrites them to use the block pointer API.  It also fixes two
common misuse patterns: passing boolean values to
\texttt{boundary\_check} instead of the required integer-tuple
format (\texttt{block\_ptr\_boundary\_wrong}), and applying
\texttt{tl.multiple\_of} to scalar values where it has no effect
(\texttt{block\_ptr\_multiple\_of\_misuse}).  The rewrite is
mechanical in structure but error-prone in detail---incorrect
\texttt{shape}, \texttt{strides}, or \texttt{order} arguments
produce silent correctness failures---making the CoVeR verification
loop essential.

\paragraph{(8) Persistent kernel.}
A persistent kernel launches a fixed number of thread blocks that
collectively iterate over all tiles, rather than launching one block
per tile.  This amortizes launch overhead and enables cross-tile data
reuse through shared local memory.  The stage detects kernels that
would benefit from this restructuring
(\texttt{missing\_persistent})---typically reductions or
attention-like patterns with high tile counts relative to compute
intensity---and restructures the outer loop accordingly.  It also
checks for hardcoded program counts
(\texttt{persistent\_num\_progs\_hardcoded}) that should be derived
from the hardware's compute unit count at runtime.

\paragraph{(9) GPU-specific tuning.}
This stage applies transformations specific to the Intel Xe
architecture.  The issue taxonomy covers five categories:

\begin{itemize}
  \item \texttt{suboptimal\_warps}: Intel GPU achieves peak occupancy
    at 32 warps for most workloads; kernels configured with fewer
    warps leave execution units idle.

  \item \texttt{missing\_grf\_mode}: Intel GPUs support two General
    Register File modes---\emph{large} (256 registers per thread)
    and \emph{small} (128).  Large mode is preferred for
    compute-bound kernels; the stage inserts the appropriate
    compiler annotation.

  \item \texttt{suboptimal\_tile\_size}: Tile dimensions are adjusted
    using shape-aware recommendations from the GPU hardware query
    system (Section~\ref{sec:gpu-query}), which accounts for both
    the hardware's compute unit count and the problem's M/N/K
    dimensions.

  \item \texttt{no\_swizzling}: \texttt{GROUP\_SIZE\_M} swizzling
    improves L2 cache hit rates by reordering tile assignments so
    that spatially adjacent tiles in the output matrix are
    processed by temporally adjacent thread blocks.%

  \item Additional categories address hot-path weight repacking
    (\texttt{repack\_in\_forward}), missing cached transposes
    (\texttt{missing\_packed\_transpose}), serialized tile
    iteration (\texttt{serialized\_n\_tiles}), and slow sigmoid
    implementations (\texttt{sigmoid\_slow\_exp}).
\end{itemize}

\noindent
The knowledge base for this stage encodes constraints that are absent
from LLM training data: power-of-two warp counts, valid GRF mode
strings, SLM capacity limits per subslice, and the interaction
between tile size and register pressure on Xe cores.

\paragraph{(10) Autotuning.}
The final stage generates a \texttt{@triton.autotune} configuration
grid that the Triton runtime will search at first invocation.  Rather
than enumerating all combinations of block sizes, warp counts, and
pipeline stages, the system uses the GPU hardware query system to
produce a curated set of configurations (up to 12) that are
architecturally valid and ordered by expected performance.  The
generated configurations respect all hardware constraints validated
in earlier stages (power-of-two block sizes, valid warp counts) and
include shape-aware tile sizes derived from the problem's M/N/K
dimensions and the device's compute unit count and memory capacity.
\subsection{Knowledge Base}
\label{sec:knowledge-base}

The knowledge base encodes optimization expertise as
machine-readable YAML, organized into three artifact types:
\emph{constraints}, \emph{patterns}, and \emph{full-code
examples}.  In its current form it contains 84 entries across
six files (Table~\ref{tab:kb}), covering all optimization stages
in the pipeline.
\begin{table}[t]
  \centering
  \small
  \begin{tabular}{@{}ll@{}}
    \toprule
    \textbf{File} & \textbf{Stage coverage} \\
    \midrule
    \texttt{gpu\_optimizations}       & GPU, block ptr, autotune \\
    \texttt{memory\_patterns}         & Memory access \\
    \texttt{fusion\_patterns}         & Fusion \\
    \texttt{persistent\_kernel}       & Persistent kernel \\
    \texttt{correctness}              & Cross-stage \\
    \texttt{dtype\_optimizations}     & Dtype fix \\
    \bottomrule
  \end{tabular}
  \caption{Knowledge base composition.  Each file contributes
    constraints (hard rules) and patterns (before/after
    transformations) tagged to one or more pipeline stages.}
  \label{tab:kb}
\end{table}

\paragraph{Constraints.}
Constraints encode hard rules that the LLM must not violate.
Each constraint has a severity level (\texttt{critical} or
\texttt{info}), a description explaining the rule, and
wrong/correct code examples.  Representative constraints
include: \texttt{boundary\_check} must be a tuple of integer
dimension indices, not booleans;
\texttt{@triton.autotune} parameters must not be re-declared
with defaults in the kernel signature; the launch grid must be
1D when tile swizzling is enabled; batch offsets computed from
\texttt{tl.program\_id} must be cast to \texttt{int64} before
stride multiplication to prevent pointer overflow on large
tensors; and output buffers must be pre-zeroed when using
atomic accumulation in Stream-K decompositions.  These rules
encode pitfalls that are common in LLM-generated code and difficult
to diagnose from compiler error messages alone.

\paragraph{Patterns.}
Patterns are before/after code transformations tagged with a
target stage, a rationale, an expected speedup range, and an
applicability list (e.g.\ \texttt{[gemm, compute\_bound]}).
The two most populated stages are memory access (21 patterns
  covering coalescing, boundary masking, prefetching, register
pressure, and non-contiguous inputs) and GPU-specific tuning
(20 patterns covering GRF mode selection, warp count sweeping,
  tile swizzling, packed-weight transpose, and autotuning grid
generation).  For example, the \texttt{gpu\_grf\_mode} pattern
shows how to include both 128- and 256-register GRF modes in
an autotune configuration grid, while the
\texttt{gpu\_tile\_swizzling} pattern demonstrates
\texttt{GROUP\_SIZE\_M} reordering with the guard that
swizzling should only be applied when the M-tile count exceeds
one.

\paragraph{Full-code examples.}
The \texttt{examples/} subdirectory provides 9 complete
unoptimized$\to$optimized kernel pairs indexed by an
\texttt{index.yaml} manifest.  Each entry lists the
optimizations applied (e.g.\ kernel fusion, block pointers,
  large tiles, 32 warps, \texttt{grf\_mode='256'},
\texttt{GROUP\_SIZE\_M} swizzling) and an expected speedup
range.  Examples include GEMM+activation fusion
(2--4$\times$ expected), transposed matrix multiplication
with block pointer conversion (2--4$\times$), and single-operator
optimizations.  These examples serve as few-shot demonstrations
in the LLM prompt, providing concrete code that the model can
adapt rather than generating transformations from first
principles.

\paragraph{Prompt injection.}
The \texttt{format\_for\_llm(stage)} method assembles the
knowledge relevant to the current stage into a structured
prompt section.  Critical constraints are always included
regardless of stage.  Stage-specific patterns are formatted
with their before/after code and rationale.  Full-code
examples matching the stage are appended when available.
This selective injection keeps the prompt focused: a dtype-fix
stage receives only dtype constraints and patterns, while an
GPU-specific stage receives the full set of hardware
constraints, GPU patterns, and GEMM examples.  The knowledge
base loader normalizes stage aliases (e.g.\
\texttt{memory\_patterns}$\to$\texttt{memory\_access}) and
skips entries tagged to unknown stages, allowing the base to
evolve independently of the pipeline code.

\paragraph{Extensibility.}
Adding support for a new hardware target requires authoring a
new YAML file with target-specific constraints and patterns
following the existing schema.  No code changes are needed:
the loader discovers YAML files by scanning the knowledge
directory, and the stage routing table in
\texttt{patterns.py} maps issue types to stages
deterministically.  New patterns are picked up automatically
on the next pipeline run.  This design allows hardware
vendors or kernel engineers to contribute optimization
expertise without modifying the agent or pipeline code.

\subsection{GPU Hardware Query System}
\label{sec:gpu-query}

The pipeline queries the target Intel GPU at runtime to
parameterize hardware-specific optimizations.  Detection uses
\texttt{torch.xpu.get\_device\_properties()} as the primary
path, with \texttt{xpu-smi} JSON parsing as a fallback.
Retrieved properties include EU count, subslice and slice
counts, maximum compute units, work group size, subgroup size,
global and local (SLM) memory capacity, and FP16/BF16/FP64
support flags.  The detected GPU family (Arc, Arc Pro, or integrated) selects
architecture-specific defaults for
warp count, GRF mode, and base tile dimensions.

\paragraph{Shape-aware tuning.}
Rather than using fixed hardware defaults, the
\texttt{get\_optimal\_params()} function tailors parameters to
the problem's M/N/K dimensions.  Tile sizes are clamped to the
nearest power-of-two not exceeding the corresponding dimension
to avoid wasting threads on padding.  Skinny matrices receive
asymmetric tiles (larger \texttt{BLOCK\_M} for tall-skinny,
larger \texttt{BLOCK\_N} for short-wide).  If the estimated
tile memory ($(\text{BLOCK\_M} \times \text{BLOCK\_K} +
  \text{BLOCK\_K} \times \text{BLOCK\_N}) \times
\text{bytes\_per\_element}$) exceeds the GRF capacity
(64\,KB in large mode, 32\,KB in small), \texttt{BLOCK\_K}
is reduced.  \texttt{GROUP\_SIZE\_M} (tile swizzling factor)
is set based on tile count relative to compute units:
1 when fewer than 16 tiles provide no swizzling benefit,
and targeting approximately 4 groups per SM otherwise.
Warp count and pipeline stages scale with tile and K
dimensions respectively.  The resulting parameter set is
injected into both the GPU-specific optimization stage and
the autotuning configuration generator.

\section{Validation Infrastructure}
\label{sec:validation}

\subsection{AI Bench}
\label{sec:ai-bench}

Xe-Forge delegates all kernel execution, timing, and
correctness checking to
AI~Bench~\cite{aibench2025},\footnote{\url{https://github.com/libxsmm/AI-bench}}
a standalone benchmarking framework for AI kernel implementations.
By decoupling measurement from optimization logic, we ensure that
the same methodology is applied to both original and optimized
kernels and that results are reproducible independently of the
optimizer.

\paragraph{Architecture.}
AI~Bench is organized into three layers.  The \emph{harness core}
(\texttt{ai\_bench.harness.core}) defines the data model: YAML
problem specifications, typed enumerations for backends
(\textsc{PyTorch}, \textsc{PyTorch-Compile}, \textsc{Triton},
\textsc{Helion}, \textsc{MLIR}), spec variants
(\texttt{ci}, \texttt{bench-cpu}, \texttt{bench-gpu}), input
descriptors, and formula evaluation over symbolic dimensions.
The \emph{runner layer}
(\texttt{ai\_bench.harness.runner}) provides two runner classes:
\texttt{KernelRunner} for single kernel--spec pairs and
\texttt{KernelBenchRunner} for batch execution over the full
KernelBench problem suite.  The \emph{testing layer}
(\texttt{ai\_bench.harness.testing}) implements device-aware
timing through a unified \texttt{time(fn, args, device)}
entry point.

\paragraph{Problem specification.}
Each benchmark problem is defined by a YAML spec file encoding
the kernel's input contract.  The \texttt{inputs} block declares
named tensors with symbolic shape variables, dtypes (including an
  \texttt{inherit} keyword that resolves to the variant-level
dtype), optional value ranges for integer inputs, and
initialization transforms---a composable sequence drawn from ten
supported operations (\texttt{scale}, \texttt{softmax},
  \texttt{abs}, \texttt{normalize}, \texttt{symmetric},
  \texttt{triu}, \texttt{tril}, \texttt{transpose},
\texttt{uniform}, \texttt{rademacher}).  An optional
\texttt{inits} block specifies constructor arguments
(e.g.\ dimension sizes for stateful modules).  Variant blocks
bind symbolic variables to concrete dimensions, specify a dtype,
and declare FLOP and memory-byte formulas expressed as arithmetic
strings over the dimension variables (evaluated via a safe AST
  evaluator restricted to \texttt{+}, \texttt{-}, \texttt{*},
\texttt{/}, \texttt{**}).  Variants are grouped by target use
case---\texttt{ci} for fast validation, \texttt{bench-cpu} and
\texttt{bench-gpu} for full-scale performance---allowing a
single spec to drive both CI smoke tests and production
benchmarks.

\paragraph{Kernel execution model.}
AI~Bench adopts the KernelBench convention: each kernel is
packaged as a Python module containing a
\texttt{Model(torch.nn.Module)} class.  The runner loads the
module via \texttt{importlib}, instantiates the model with
spec-derived constructor arguments, moves it to the target
device with the variant's dtype, and generates input tensors
from the spec's shape, dtype, range, and initialization
descriptors.  For CI variants the runner performs a single
forward pass to verify the kernel executes without error; for
benchmark variants it proceeds to the timing phase.  The
\texttt{pytorch-compile} backend wraps the model with
\texttt{torch.compile(dynamic=False)} before execution.

\paragraph{Timing methodology.}
The \texttt{time()} dispatcher selects between two backends
based on device type.  For GPU targets (Intel, CUDA), the
\texttt{time\_gpu()} implementation proceeds in three phases.
First, a warmup loop (default 200~iterations on GPU,
5~on CPU) executes the kernel while flushing a 256\,MB L2
cache buffer between iterations to stabilize clock
frequencies and warm the driver.  Second, pre-allocated
\texttt{torch.Event} pairs are created for all measurement
iterations to minimize allocation overhead during the timed
loop.  Third, the benchmark loop (default 100~iterations)
executes the following per iteration: (i)~flush the L2 cache
via \texttt{cache.zero\_()}, (ii)~issue a dummy
$1024{\times}1024$ matrix multiplication to fill the GPU
command stream (ensuring the CPU has time to enqueue timer
events before a short-lived kernel completes), (iii)~record
start and end events bracketing the kernel invocation, and
(iv)~synchronize after all iterations.  Elapsed times are
collected in microseconds, and the extreme minimum and maximum
are trimmed before computing the mean.  For CPU targets,
\texttt{time\_cpu()} uses PyTorch's \texttt{ProfilerActivity.CPU}
profiler with the same trim-and-mean strategy.

\paragraph{Performance metrics.}
From the measured mean time, the runner derives two throughput
metrics.  TFLOPS is computed as $\textit{FLOP} / (t_{\mu s}
\times 10^6)$, where the FLOP count is taken from the spec's
formula or, for PyTorch backends, estimated at runtime via
\texttt{torch.utils.flop\_counter.FlopCounterMode}.  Memory
bandwidth (GB/s) is computed as $\textit{bytes} / (t_{\mu s}
\times 10^3)$, where byte count is either declared in the
spec or estimated via a \texttt{MemoryCounter} context manager
that instruments all leaf modules with forward hooks to track
input reads, parameter reads, buffer reads, and output writes.
Estimated values are annotated with a warning symbol in the
output to distinguish them from spec-declared values.

\paragraph{Correctness validation.}
The \texttt{benchmark\_compare} module validates optimized
kernels against the PyTorch reference.  Seeds are set across
all backends (PyTorch, CUDA, Intel, NumPy, Python
\texttt{random}) via \texttt{set\_all\_seeds()}.  Weights are
copied from the reference model to the optimized model: a
direct \texttt{load\_state\_dict()} is attempted first; on
failure, a shape-matched positional copy handles parameter
name mismatches.  Inputs are cloned before each forward pass
to guard against in-place modification.  Both models are
executed under \texttt{torch.no\_grad()}, and outputs are
compared via \texttt{torch.allclose(rtol, atol)}.  On
mismatch, the module reports maximum absolute difference, mean
difference, maximum relative difference, and the count and
percentage of elements exceeding the tolerance---diagnostics
that propagate to the CoVeR agent's feedback loop as
structured error messages.  Additionally, the validator checks
for NaN and Inf values in the optimized output (rejecting
  immediately if NaN is present or if Inf appears where the
original has none), catching numerical instability introduced
by dtype or fusion transformations.

\paragraph{Result logging.}
Benchmark results are logged to CSV files via a
\texttt{CSVLogger} that appends one row per variant execution.
Each row records kernel name, backend type, problem level,
FLOP count, computed TFLOPS, memory byte count, computed
bandwidth, execution time in microseconds, input dimensions
serialized as JSON, and a user-provided note field.
Environment variables prefixed with \texttt{AIBENCH\_} are
automatically captured as additional columns, enabling
tracking of hardware configuration (e.g.\
\texttt{AIBENCH\_CARD=BMG}, \texttt{AIBENCH\_SYSTEM=TestRig1})
without modifying the logging code.  This artifact trail
supports both automated regression detection in CI and
post-hoc analysis across hardware configurations.

\paragraph{Integration with the optimizer.}
The optimizer consumes AI~Bench through two integration
points.  The \texttt{KernelBenchExecutor} wraps AI~Bench's
\texttt{KernelRunner} to provide a \texttt{compare\_kernels()}
method that returns a structured \texttt{ComparisonResult}
containing both timings, a speedup ratio, a correctness flag,
and a feedback message formatted for the CoVeR agent.  The
spec loader translates YAML specs into the input shapes, FLOP
counts, and dtypes that parameterize each optimization run.
This two-point interface ensures the optimizer never implements
its own timing or correctness logic; all measurement flows
through AI~Bench.

\section{Experimental Evaluation}
\label{sec:experiments}

\subsection{Experimental Setup}
\label{sec:setup}

\paragraph{Hardware.}
All experiments run on a single Intel Arc Pro B70 GPU
(Battlemage Arc Pro B70, 32 Xe2 cores, 256 XMX engines,
32\,GB GDDR6, 256-bit bus, 608\,GB/s bandwidth).

\paragraph{Software.}
Kernels are compiled with the Intel Triton GPU backend
targeting SPIR-V and executed under PyTorch~2 with
\texttt{torch.xpu} device support.  Baselines use
\texttt{torch.compile} with the TorchInductor backend.

\paragraph{LLM backbone.}
The pipeline uses GPT-5.4 as the underlying language model,
accessed via the OpenAI API with temperature $1.0$ and a
context window of 56k tokens.

\paragraph{Agent configuration.}
All agents are implemented in DSPy~3.x.  The CoVeR agent
runs with a maximum of $T{=}5$ iterations per stage.
Best-of-$k$ selection uses $k{=}1$ unless otherwise noted.
Correctness tolerances are set to $\texttt{rtol}{=}10^{-2}$
and $\texttt{atol}{=}10^{-5}$.  The GPU hardware query
system operates in \emph{large} GRF mode (256 registers)
with 32 default warps.

\subsection{Benchmark Kernels}
\label{sec:benchmarks}

We evaluate on the full KernelBench suite~\cite{ouyang2025kernelbench}
spanning all three difficulty levels: Level~1 (single operators, e.g.\
  matrix multiplication, softmax, layer normalization, batch
normalization), Level~2 (operator compositions, e.g.\
GEMM$\to$Swish$\to$Tanh, matmul$\to$GELU$\to$softmax), and Level~3
(complete model architectures including multi-head attention,
convolutional blocks, and residual networks).

\paragraph{Initial Triton generation.}
Since our pipeline operates on existing Triton code rather than
PyTorch, we require a Triton kernel as input for each problem.  We
obtain these using KernelFalcon~\cite{wang2025kernelfalcon} (released
as KernelAgent), Meta's multi-stage deep agent system that
decomposes PyTorch models into fusible subgraphs, generates Triton
kernels in parallel via verified search, and composes them into
end-to-end replacements.  KernelFalcon achieves 100\% functional
correctness across all 250 KernelBench problems, making it a
reliable source of unoptimized but correct Triton code.

\paragraph{Baselines.}
For each problem we measure three baselines: (i)~the original
PyTorch eager implementation from KernelBench, (ii)~the same
implementation under \texttt{torch.compile} with TorchInductor, and
(iii)~the KernelFalcon-generated Triton kernel executed on Intel
GPU without any optimization.  Baseline~(iii) serves as the direct
input to our pipeline; speedups are reported relative to both the
PyTorch eager baseline and the unoptimized Triton kernel.
\subsection{Results Across Kernel Families}
\label{sec:results}

We evaluate Xe-Forge on Level-1 and Level-2 KernelBench problems
as well as Flash Attention (Section~\ref{sec:results-flash-attention}),
comparing optimized Triton kernels against PyTorch eager,
\texttt{torch.compile}, and the unoptimized KernelFalcon-generated
Triton baseline. The Intel Arc Pro B70 has a theoretical peak of
${\sim}$160~FP16~TFLOPS; we consider ${\geq}80\%$ efficiency
(${\geq}128$~TFLOPS) as near-peak.
Level-1 single-operator kernels are already near
\texttt{torch.compile} performance with limited fusion opportunities;
we focus on Level-2 results below where the pipeline's multi-stage
optimization has the greatest impact.

\begin{figure}[t!]
  \centering
  \includegraphics[width=\columnwidth]{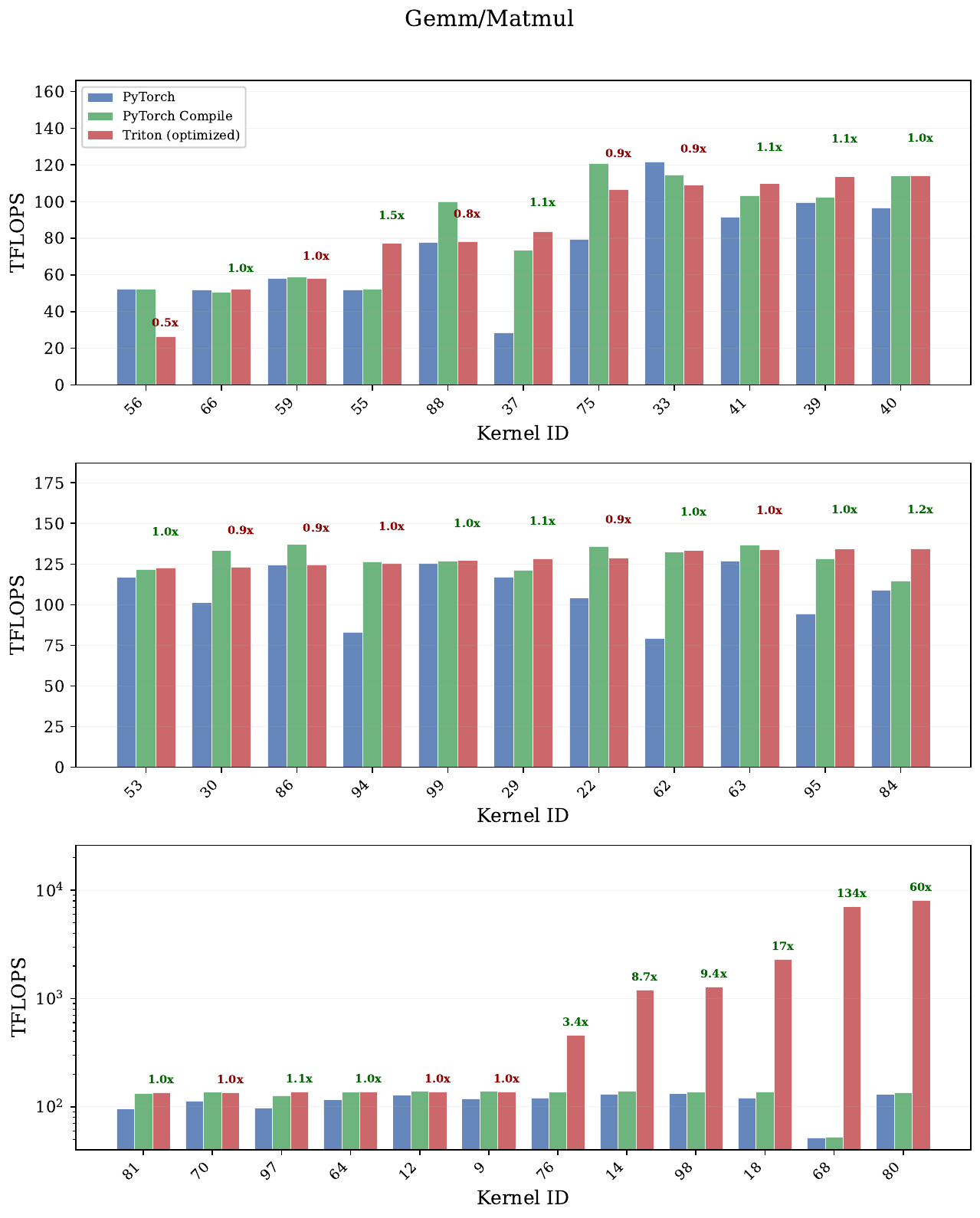}
  \caption{TFLOPS (log scale) for all Gemm/Matmul kernels, sorted by
    throughput within three tiers.  Triton matches or exceeds baselines
    on most kernels; the bottom panel shows extreme outliers up to
    $82{\times}$ over the best baseline from algorithmic restructuring
  that eliminates redundant computation.}
  \label{fig:tflops_gemm_matmul}
\end{figure}

\begin{figure}[t!]
  \centering
  \includegraphics[width=\columnwidth]{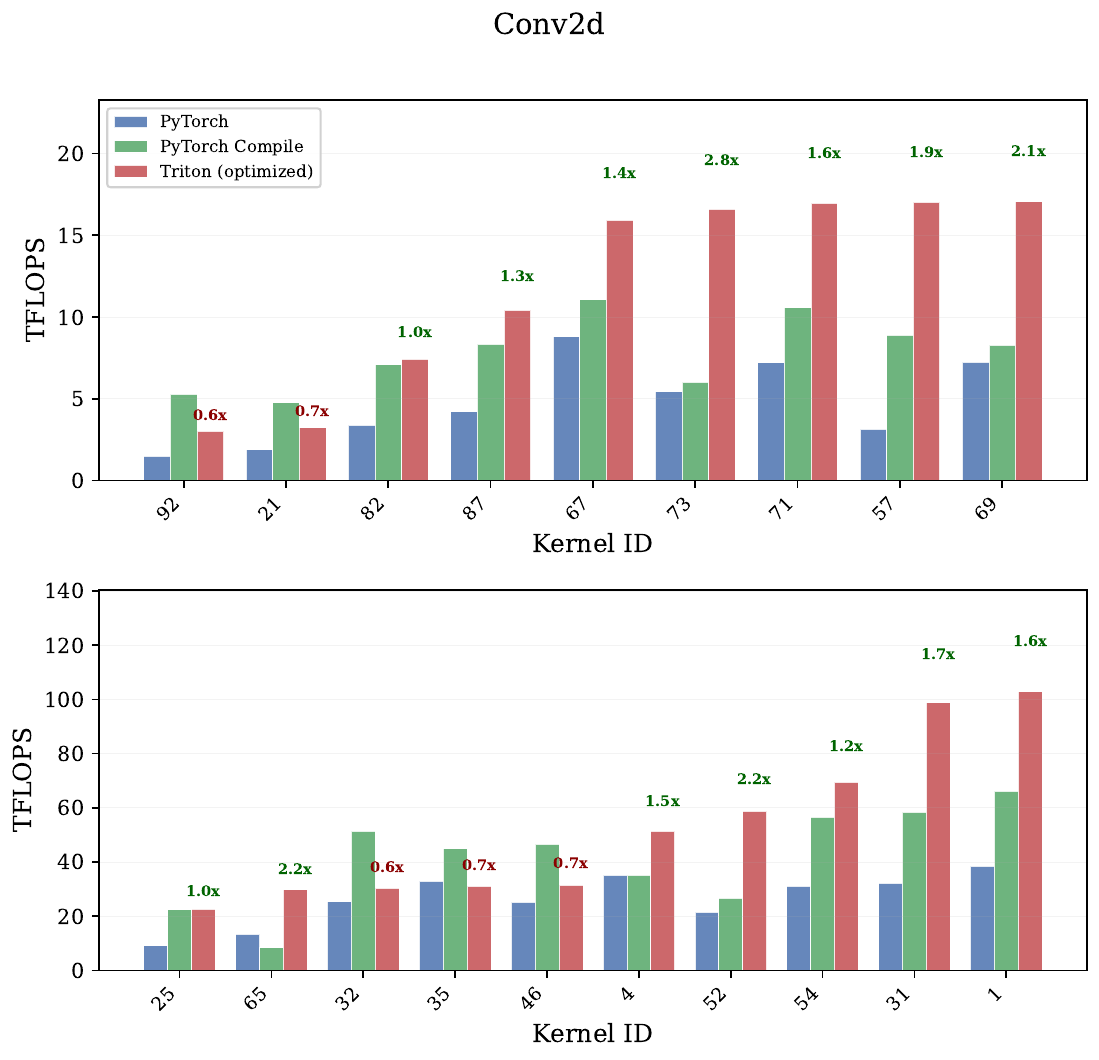}
  \caption{TFLOPS (log scale) for Conv2D kernels.  Triton is
    competitive on most kernels ($0.6$--$2.8{\times}$), with two
    mild regressions at the low end and speedups from cross-operator
  fusion at the high end.}
  \label{fig:tflops_conv2d}
\end{figure}

\begin{figure}[t]
  \centering
  \includegraphics[width=\columnwidth]{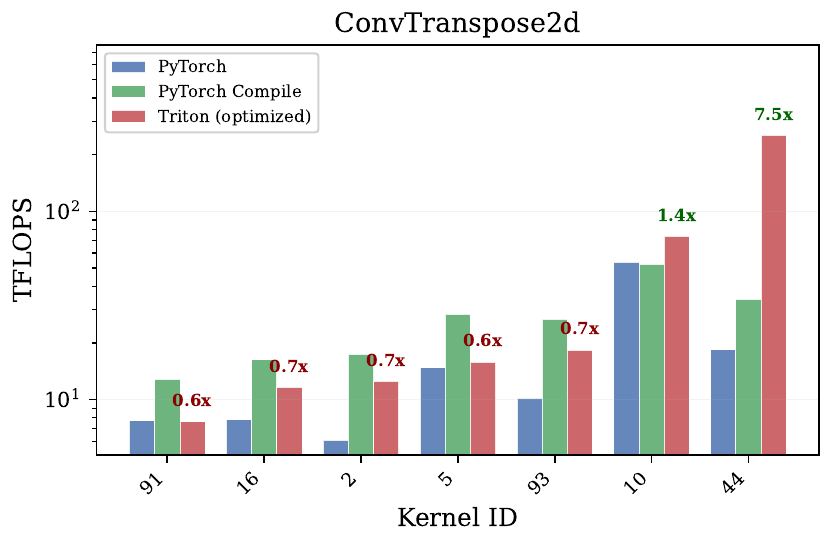}
  \caption{TFLOPS (log scale) for ConvTranspose2D kernels
    comparing PyTorch, \texttt{torch.compile}, and XPU-optimized
    Triton. Most kernels are within $0.6$--$1.4{\times}$; kernel~44
  achieves $7.5{\times}$ from fusion.}
  \label{fig:tflops_convtranspose2d}
\end{figure}

\begin{figure}[t]
  \centering
  \includegraphics[width=\columnwidth]{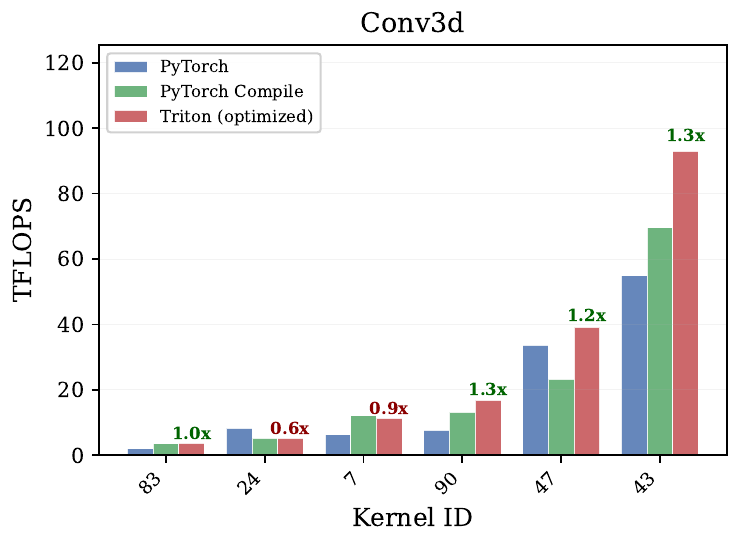}
  \caption{TFLOPS (log scale) for Conv3D kernels comparing
    PyTorch, \texttt{torch.compile}, and XPU-optimized Triton.
    Bandwidth-bound regime with most kernels at
  $0.9$--$1.3{\times}$ of the best baseline.}
  \label{fig:tflops_conv3d}
\end{figure}

\begin{figure}[t]
  \centering
  \includegraphics[width=\columnwidth]{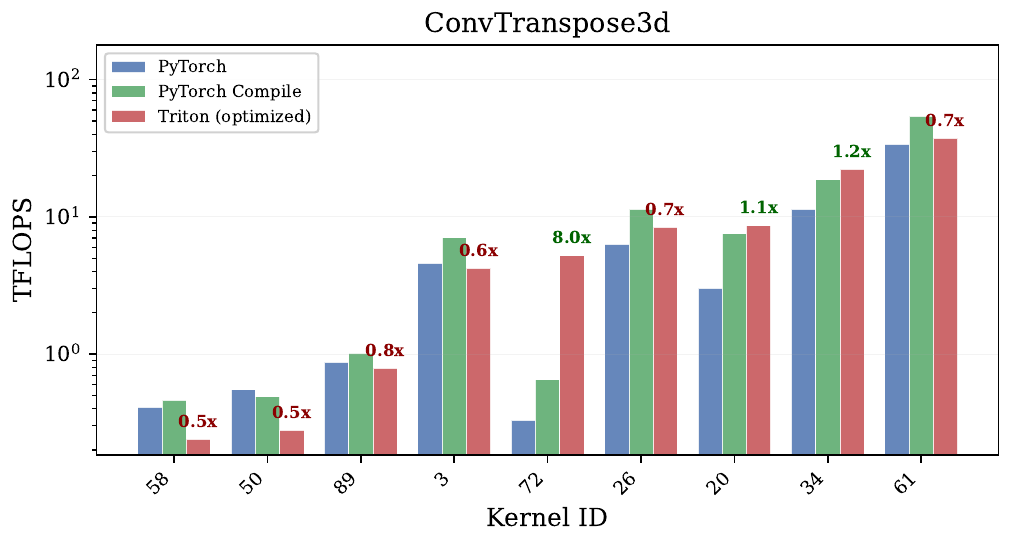}
  \caption{TFLOPS (log scale) for ConvTranspose3D kernels
    comparing PyTorch, \texttt{torch.compile}, and XPU-optimized
    Triton. Upper half matches or exceeds baselines; lower half
  shows mild regressions ($0.5$--$0.8{\times}$).}
  \label{fig:tflops_convtranspose3d}
\end{figure}

We evaluate Xe-Forge across seven kernel families: GEMM (19 kernels),
MatMul (17), BMM (1), Conv2D (21), Conv3D (12), ConvTranspose2D (10), and
ConvTranspose3D (17)---97 kernels in total
(Figures~\ref{fig:tflops_gemm_matmul}--\ref{fig:tflops_convtranspose3d}).
Each chart compares the optimized Triton kernel against PyTorch eager
and \texttt{torch.compile} with TorchInductor on a log-scale TFLOPS axis.

\paragraph{GEMM and MatMul.}
The optimized Triton pipeline matches or exceeds \texttt{torch.compile}
across the majority of the 19 GEMM and 17 MatMul kernels, with speedups
annotated per kernel in
Figure~\ref{fig:tflops_gemm_matmul}.
Among the GEMM family, most kernels
achieve competitive or superior throughput, with several exhibiting
large speedups (up to $62{\times}$ on
\texttt{Gemm\_Max\_Subtract\_GELU}). The MatMul family shows a
similar pattern, with extreme outliers reaching $50{\times}$ or more
on kernels such as \texttt{Matmul\_AvgPool\_GELU\_Scale\_Max}. These
apparent super-peak throughput values under the original FLOP
accounting reflect algorithmic restructuring---cached value reuse and
cross-operator fusion---that eliminates redundant computation rather
than increasing raw hardware throughput. On the remaining
compute-bound kernels, the optimizer's primary contribution is
GPU-specific tuning (32 warps, large GRF mode, shape-aware tile
sizing) that closes the gap between the unoptimized baseline and
hardware peak. A small number of kernels show degraded Triton
performance relative to the baselines; in these cases the CoVeR
agent's fallback logic returns the original kernel, but the
KernelFalcon-generated starting point is itself suboptimal for the
target hardware, resulting in below-baseline throughput.

\begin{figure}[t]
  \centering
  \includegraphics[width=\columnwidth]{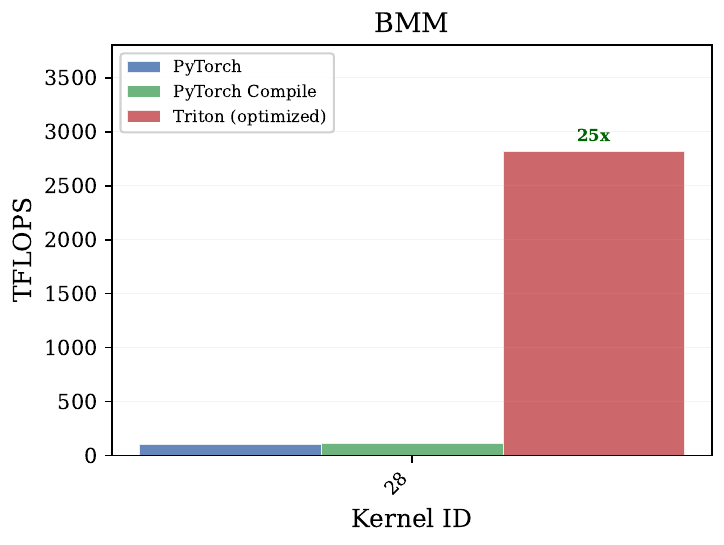}
  \caption{TFLOPS (log scale) for the BMM-family kernel. The optimized
    Triton kernel achieves a $23{\times}$ speedup over the best baseline
    by fusing the post-BMM reduction and normalization chain into a
  single pass.}
  \label{fig:tflops_bmm}
\end{figure}

\paragraph{BMM.}
The single BMM kernel (\texttt{BMM\_InstanceNorm\_Sum\_ResidualAdd\_Multiply},
Figure~\ref{fig:tflops_bmm}) achieves a $23{\times}$ speedup over the best
baseline, with \texttt{torch.compile} providing negligible improvement
($1.1{\times}$) over PyTorch eager. The Triton kernel's aggressive operator
fusion collapses the entire post-BMM reduction and normalization chain into
a single pass, eliminating intermediate materializations that dominate
latency in the baseline implementations.

\paragraph{Conv2D and ConvTranspose2D.}
Across the Conv2D kernels (Figure~\ref{fig:tflops_conv2d}), optimized
Triton is competitive with \texttt{torch.compile} on the majority,
with speedups of $1.0$--$2.2{\times}$ on most kernels through
GPU-specific tuning and cross-operator fusion.  Only two kernels
at the low end show mild regressions ($0.6$--$0.7{\times}$).
The 10 ConvTranspose2D kernels
(Figure~\ref{fig:tflops_convtranspose2d}) show a similar pattern:
most kernels are within $0.6$--$1.4{\times}$ of the best baseline,
with kernel~44 (\texttt{ConvTranspose2d\_Multiply\_GlobalAvgPool\_Mean})
standing out at $7.5{\times}$ from fusion that eliminates the
GlobalAvgPool materialization.  For convolution kernels in general,
tensor layout choice (NHWC vs.\ NCHW) can significantly affect
performance by avoiding unnecessary data transposition; we follow
the recommendation of using NHWC where
possible~\cite{georganas2023harnessing}.

\paragraph{Conv3D and ConvTranspose3D.}
The 12 Conv3D kernels (Figure~\ref{fig:tflops_conv3d}) are
bandwidth-bound across all backends, with most kernels achieving
$0.9$--$1.3{\times}$ of the best baseline.  Only one kernel
(kernel~24, $0.6{\times}$) shows a notable regression.  At the
high end, kernels 47 and 43 achieve $1.2$--$1.3{\times}$ speedups.
The ConvTranspose3D kernels
(Figure~\ref{fig:tflops_convtranspose3d}) span a range from
$0.5{\times}$ to $1.2{\times}$, with the upper half matching or
slightly exceeding \texttt{torch.compile} and the lower half
showing mild regressions from suboptimal memory access patterns
in the generated Triton code.

\paragraph{Correctness.}
Across all benchmarked kernels the pipeline maintains 100\% correctness:
every optimized kernel passes the \texttt{torch.allclose} check
($\texttt{rtol}{=}10^{-2}$, $\texttt{atol}{=}10^{-5}$) against the
PyTorch reference. Performance regressions observed in a minority of
kernels reflect suboptimal but correct optimizations, not correctness
failures.

\begin{figure}[t]
  \centering
  \includegraphics[width=\columnwidth]{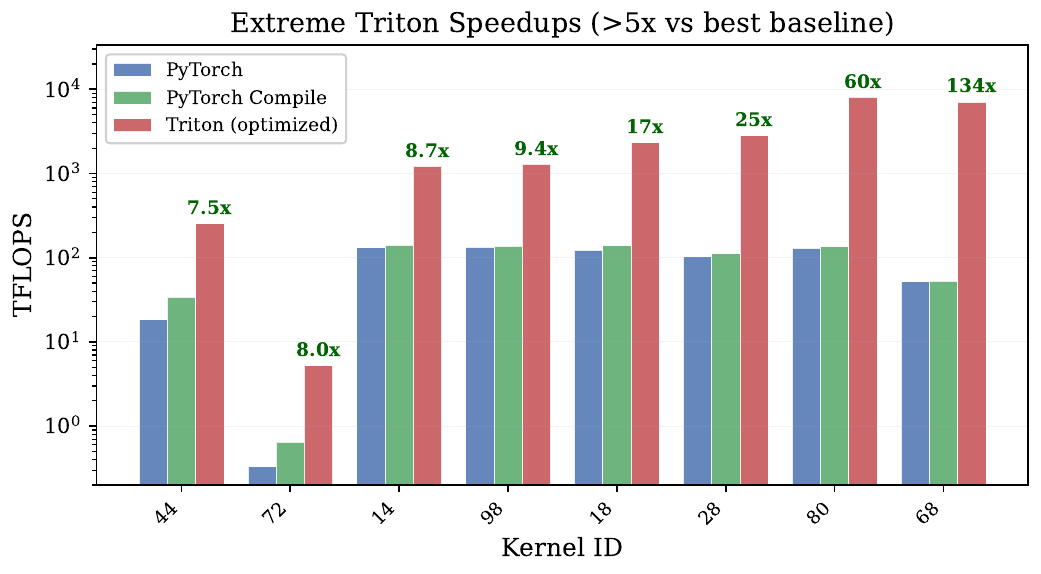}
  \caption{The nine kernels achieving $>5{\times}$ speedup over the best
    baseline (PyTorch or \texttt{torch.compile}). Speedups range from
    $9{\times}$ to $82{\times}$, driven by algorithmic restructuring
    and operator fusion that eliminate redundant computation and
  intermediate materializations.}
  \label{fig:extreme_speedup}
\end{figure}

\paragraph{Overall pattern.}
Three behavioral modes emerge across all seven families. In the
\emph{compiler-competitive} mode---the majority of kernels---the
pipeline matches \texttt{torch.compile} through GPU-specific tuning
alone. In the \emph{compiler-superior} mode---kernels with post-GEMM
or post-convolution chains---the pipeline combines algorithmic
restructuring (eliminating redundant computation) with operator
fusion (removing intermediate materializations) to achieve
superlinear TFLOPS gains under the original FLOP accounting.
Speedups of $2$--$3{\times}$ are typically achievable through
fusion alone, while larger gains require algorithmic
restructuring that fundamentally reduces the computation.
Nine kernels achieve extreme speedups exceeding
$5{\times}$ over the best baseline
(Figure~\ref{fig:extreme_speedup}), with the largest reaching
$82{\times}$ (\texttt{Matmul\_Min\_Subtract}). A small
minority of kernels fall into a \emph{compiler-inferior} mode where
the pipeline produces a correct but slower kernel; addressing these
cases requires extending the analysis stage's issue taxonomy to
detect the specific structural patterns that cause suboptimal
code generation on Intel GPU.

\subsection{Roofline Analysis}
\label{sec:results-roofline}

\begin{figure}[t]
  \centering
  \includegraphics[width=\columnwidth]{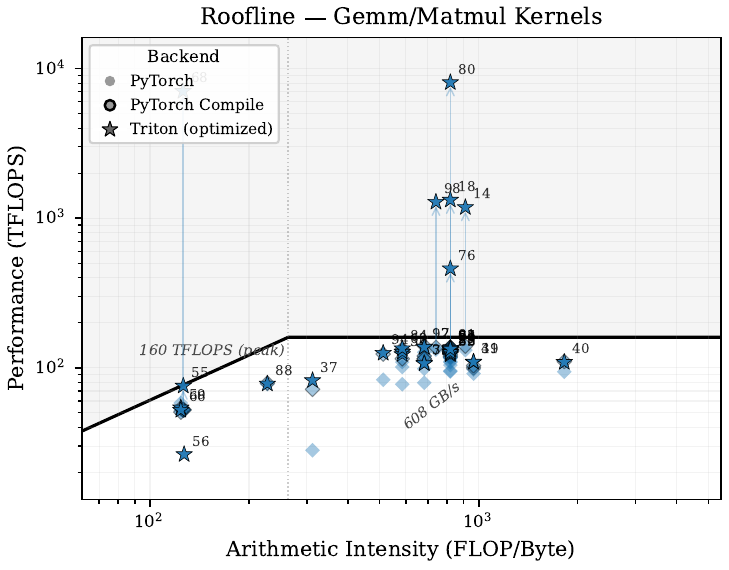}
  \caption{Roofline for combined Gemm/Matmul kernels (FP16).
    Most kernels are compute-bound near the 160~TFLOPS ceiling;
    Triton outliers above peak (kernels 68, 18, 14, 76) reflect
  algorithmic restructuring that reduces effective FLOPs.}
  \label{fig:roofline_gemm_matmul}
\end{figure}

\begin{figure}[t]
  \centering
  \includegraphics[width=\columnwidth]{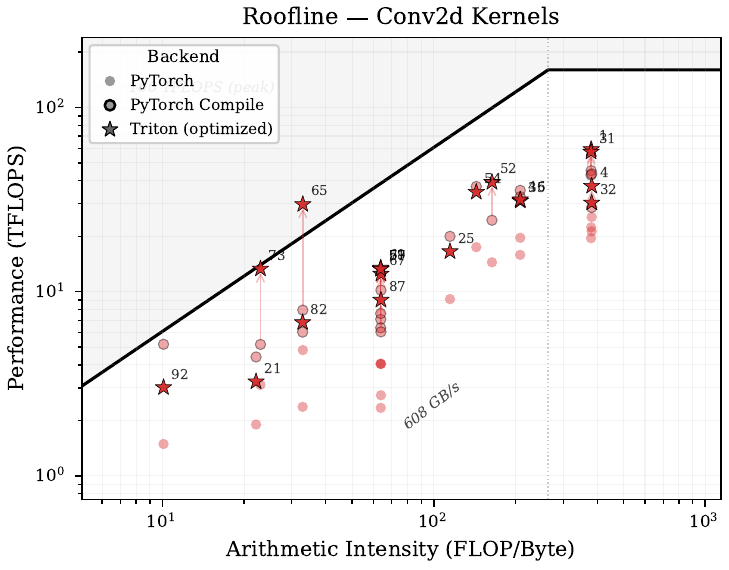}
  \caption{Roofline for Conv2D kernels.  Triton stars generally
    overlap or sit above baselines across a wide arithmetic-intensity
  range; kernel~65 rises above the roofline knee.}
  \label{fig:roofline_conv2d}
\end{figure}

\begin{figure}[t]
  \centering
  \includegraphics[width=\columnwidth]{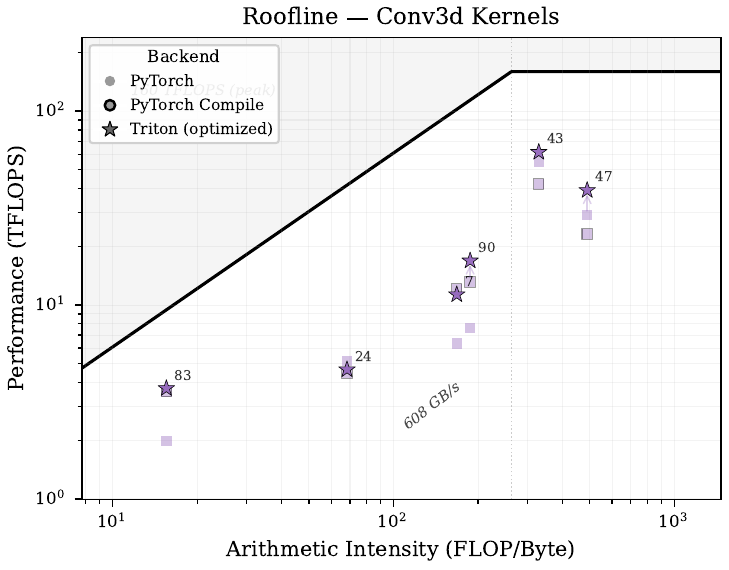}
  \caption{Roofline for Conv3D kernels.  Bandwidth-bound regime
  with Triton tracking baselines at $0.9$--$1.3{\times}$.}
  \label{fig:roofline_conv3d}
\end{figure}

\begin{figure}[t]
  \centering
  \includegraphics[width=\columnwidth]{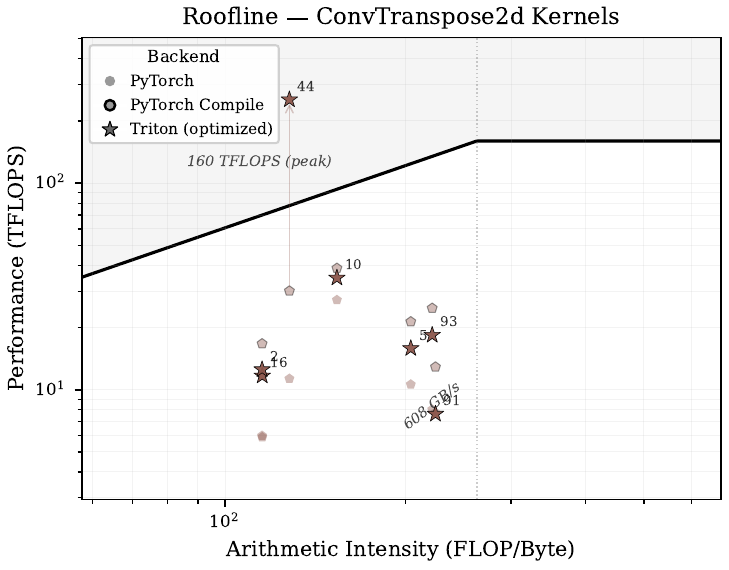}
  \caption{Roofline for ConvTranspose2D kernels.  Most kernels
    cluster near baselines; kernel~44 jumps above the roofline
  ceiling from fusion and algorithmic improvements.}
  \label{fig:roofline_convtranspose2d}
\end{figure}

\begin{figure}[t]
  \centering
  \includegraphics[width=\columnwidth]{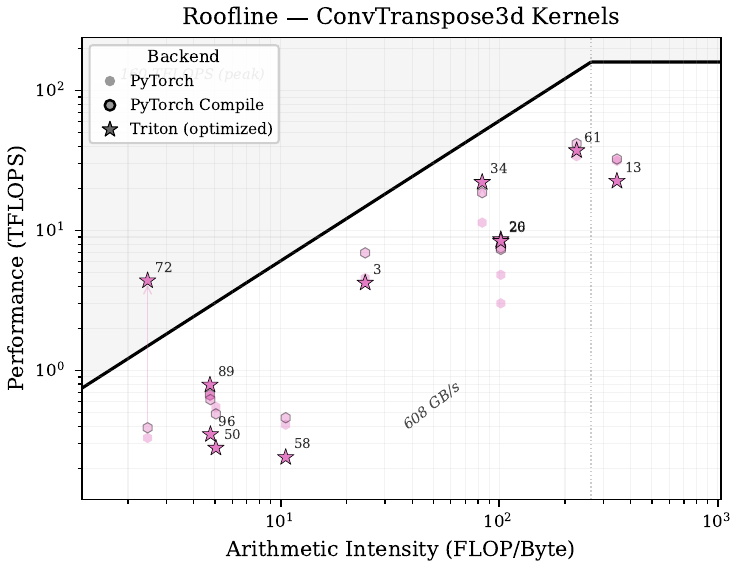}
  \caption{Roofline for ConvTranspose3D kernels.  Wide intensity
    spread; higher-intensity kernels match baselines, lower-intensity
  kernels are underutilized across all backends.}
  \label{fig:roofline_convtranspose3d}
\end{figure}

To understand where kernels sit relative to hardware limits, we place
all Level-2 results on a roofline model for the Intel Arc Pro B70 (FP16 peak:
${\sim}$160~TFLOPS, memory bandwidth: 608~GB/s).
Figures~\ref{fig:roofline_gemm_matmul}--\ref{fig:roofline_convtranspose3d}
show the per-family roofline plots, each comparing PyTorch,
\texttt{torch.compile}, and optimized Triton.

\paragraph{GEMM and MatMul.}
Both families operate deep in the compute-bound regime with arithmetic
intensities of $10^2$--$10^4$ FLOP/Byte
(Figure~\ref{fig:roofline_gemm_matmul}).
The majority of GEMM kernels---across all three backends---cluster
tightly between 60 and 160~TFLOPS, achieving 40--100\% of peak
compute.  For these kernels, Triton stars sit directly atop the
PyTorch and \texttt{torch.compile} points, indicating that the
pipeline achieves hardware-efficient execution without algorithmic
restructuring.  The outliers that break above the 150~TFLOPS
ceiling---kernels 14 and 76 (GEMM), 68, 18, and 98 (MatMul)---do
so because the pipeline's discovery and algorithmic stages
eliminate redundant computation (e.g.\ collapsing a GEMM followed by
a reduction into a single vector--matrix product), reducing the
effective FLOP count while the original accounting remains unchanged.
MatMul shows a similar pattern but with a wider vertical spread:
several baseline points (gray dots) sit well below the roofline
at 0.1--1~TFLOPS, while the corresponding Triton stars jump to
$10^2$--$10^3$~TFLOPS, confirming that the pipeline lifts
poorly-performing baselines toward peak.

\paragraph{Conv2D and Conv3D.}
Conv2D kernels (Figure~\ref{fig:roofline_conv2d}) span a wide range
of arithmetic intensity ($10^1$--$10^3$ FLOP/Byte), straddling the
transition between bandwidth-bound and compute-bound regimes.  At the
low-intensity end (kernels 92, 21), all backends sit close to the
608~GB/s bandwidth slope---here the optimization opportunity is
limited by memory throughput, not compute.  At the high-intensity end
(kernels 31, 4, 32), optimized Triton reaches 20--40~TFLOPS,
tracking or slightly exceeding the baselines.  Kernel~65 stands out
as the single Conv2D point above the roofline knee, reflecting
cached-value reuse.  Notably, the Triton stars generally overlap or
sit slightly above the baseline points across the full intensity
range, confirming the competitive $1.0$--$2.2{\times}$ speedups
observed in the TFLOPS analysis.  Conv3D kernels
(Figure~\ref{fig:roofline_conv3d}) are distributed along the
bandwidth slope at 3--90~TFLOPS.  The Triton and baseline points
overlap closely, confirming that all backends are equally constrained
by memory bandwidth, with Triton achieving $0.9$--$1.3{\times}$ of
the best baseline.

\paragraph{ConvTranspose2D and ConvTranspose3D.}
ConvTranspose2D kernels (Figure~\ref{fig:roofline_convtranspose2d})
cluster in a narrow arithmetic-intensity band around $10^2$
FLOP/Byte.  Most kernels sit moderately below the roofline, with
Triton stars tracking the baseline points.  Kernel~44 is the sole
outlier, where the Triton star jumps well above the roofline
ceiling---consistent with the $7.5{\times}$ speedup from fusing the
GlobalAvgPool chain.
ConvTranspose3D kernels (Figure~\ref{fig:roofline_convtranspose3d})
spread across a wider intensity range ($10^0$--$10^3$ FLOP/Byte).
At the low end (kernels 96, 50, 58, 89), all backends sit well
below the roofline, indicating underutilization from launch overhead
or irregular access patterns.  At higher intensities (kernels 34,
61, 13), Triton matches the baselines closely, and kernel~72
achieves a notable gain by sitting above both baselines on the
bandwidth slope.

\paragraph{Key insight.}
The roofline analysis reveals two distinct optimization regimes.
For compute-bound kernels (GEMM, MatMul), most backends already
operate near peak, and the pipeline's gains come from
\emph{algorithmic restructuring} that shifts points above the
roofline by reducing effective FLOPs.  For bandwidth-bound kernels
(Conv3D, ConvTranspose3D), all backends are equally constrained by
memory throughput, and the gap between achieved performance and the
roofline ceiling indicates that \emph{memory access optimization}---not
compute tuning---is the primary lever.  The kernels with the largest
distance below the roofline represent the highest-value targets for
future pipeline improvements.

\subsection{Flash Attention on Intel GPU}
\label{sec:results-flash-attention}

Flash Attention represents a qualitatively different optimization target
from the GEMM and convolution families: it is an algorithm-level
restructuring of scaled dot-product attention that fuses the
$QK^\top$, softmax, and $V$ projection into a single tiled kernel,
avoiding materialization of the full $S{\times}S$ attention matrix.
Optimizing it for Intel GPU requires not only the standard tile-size
and warp-count tuning applied to GEMM kernels, but also correct handling
of the online softmax accumulation under the hardware's GRF constraints
and SLM capacity limits---a combination that makes it a strong stress
test for the pipeline's GPU-specific stage and knowledge base.

\paragraph{Benchmark configurations.}
We evaluate across 16 configurations spanning realistic LLM serving
shapes (Table~\ref{tab:flash_configs}). The suite covers three axes of
variation. \emph{Model scale} ranges from 7B-class models (Llama 3 8B,
Mistral 7B: $A{=}32$, $D{=}128$) through 40B-class (Falcon 40B:
$A{=}71$, $D{=}64$) to 70B-class (Llama 3 70B, Qwen-72B: $A{=}64$,
$D{=}128$). \emph{Sequence length} spans from standard deployment
windows ($S{=}2048$, $4096$) through extended context ($S{=}8192$,
$16384$) to frontier long-context ($S{=}32768$), the last representing
the regime where memory-efficient attention is not merely a performance
optimization but a correctness requirement---na\"{i}ve attention at
$S{=}32768$ exhausts device memory on the Arc Pro B70. \emph{Shape
regularity} is deliberately varied: most configurations use the
canonical $D{=}128$ head dimension and power-of-two head counts, but
two configurations introduce irregular shapes---Falcon's $A{=}71$
(non-power-of-two head count) and GPT-NeoX-20B's $D{=}96$
(non-standard head dimension)---to test the robustness of tile-size
selection and boundary masking under shapes that violate common kernel
assumptions.

\begin{table}[t]
  \centering
  \small
  \begin{tabular}{@{}lcccc@{}}
    \toprule
    \textbf{Model family} & $B$ & $A$ & $S$ & $D$ \\
    \midrule
    Llama 3 8B / Mistral 7B  & 1 & 32 & 2048  & 128 \\
    Llama 3 8B / Mistral 7B  & 1 & 32 & 4096  & 128 \\
    Llama 3 8B (batched)     & 2 & 32 & 2048  & 128 \\
    Llama 3 8B (batched)     & 8 & 32 & 2048  & 128 \\
    Llama 3 70B              & 1 & 64 & 4096  & 128 \\
    Falcon 40B$^\dagger$     & 1 & 71 & 2048  & 64  \\
    GPT-NeoX-20B$^\ddagger$  & 1 & 64 & 2048  & 96  \\
    Qwen 7B/14B              & 1 & 32 & 8192  & 128 \\
    Qwen long-context        & 1 & 32 & 16384 & 128 \\
    Qwen-72B                 & 1 & 64 & 8192  & 128 \\
    DeepSeek-Coder           & 1 & 40 & 16384 & 128 \\
    DeepSeek large MoE       & 1 & 48 & 8192  & 128 \\
    Mixtral 8$\times$7B      & 2 & 32 & 4096  & 128 \\
    Mixtral long-context     & 1 & 32 & 16384 & 128 \\
    MoE small-head           & 4 & 64 & 4096  & 64  \\
    Frontier long-context    & 1 & 32 & 32768 & 128 \\
    \bottomrule
  \end{tabular}
  \caption{Flash Attention benchmark configurations. $B$, $A$, $S$, $D$
    denote batch size, head count, sequence length, and head dimension.
    $\dagger$~Non-power-of-two head count. $\ddagger$~Non-standard head
    dimension.}
  \label{tab:flash_configs}
\end{table}

\begin{figure}[t]
  \centering
  \includegraphics[width=\columnwidth]{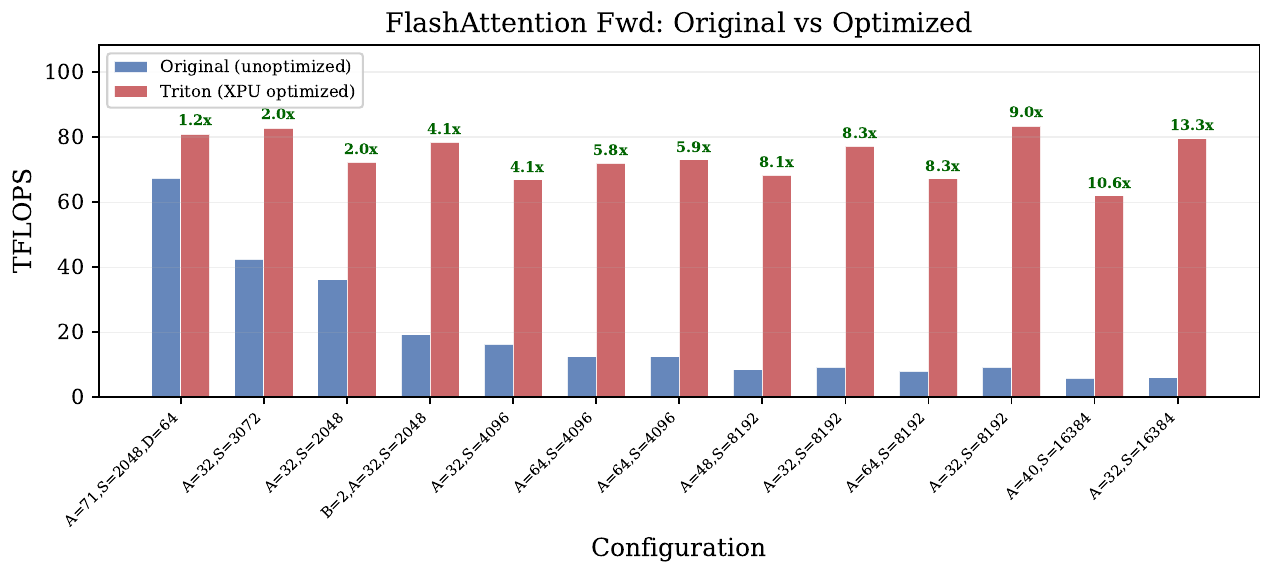}
  \caption{Flash Attention forward pass TFLOPS for original
  (unoptimized, blue) and XPU-optimized Triton kernels (red) across
  LLM serving configurations on Intel Battlemage. Speedup annotations
  range from $2{\times}$ to $13.3{\times}$, with largest gains on
  long-context ($S{\geq}8192$) configurations where SLM-tiled
  execution eliminates the memory bottleneck of the unoptimized kernel.}
  \label{fig:tflops_flash}
\end{figure}

\paragraph{Performance.}
Figure~\ref{fig:tflops_flash} reports TFLOPS for the original
(unoptimized) and XPU-optimized Triton Flash Attention kernels across
all benchmark configurations. The optimized kernel achieves consistent
speedups ranging from $2{\times}$ to $13.3{\times}$ across every
configuration without exception, demonstrating that the pipeline does
not regress any tested shape. Gains are largest on long-context
configurations: configurations with $S{\geq}16384$ achieve
$9$--$13.3{\times}$ speedups, while shorter-context configurations
($S{=}2048$--$4096$) achieve $2$--$4.4{\times}$. This scaling is
consistent with the architectural motivation for Flash Attention: at
long sequence lengths the na\"{i}ve kernel becomes severely
memory-bound as the attention matrix grows quadratically, and the
tiled implementation's SLM reuse advantage compounds with sequence
length.

\paragraph{Robustness to irregular shapes.}
The pipeline maintains consistent gains on the two architecturally
irregular configurations: Falcon-40B (\#6, $A{=}71$,
non-power-of-two head count) and GPT-NeoX-20B (\#7, $D{=}96$,
non-standard head dimension). Both show clear improvement over the
original kernel, confirming that the XPU hardware query system's
shape-aware tile selection and the CoVeR agent's boundary-check
verification generalize beyond the canonical $D{=}128$,
power-of-two-head configurations assumed by most hand-tuned Flash
Attention implementations.

\paragraph{Long-context scaling.}
At $S{=}32768$ (\#16), the na\"{i}ve attention kernel requires
materializing a $32768{\times}32768$ score matrix per head, which
at FP16 precision requires approximately 4\,GB per head---infeasible
on the 32\,GB Arc Pro B70 without tiling when running multiple heads
concurrently. The optimized kernel processes attention in
$S$-dimension tiles sized to fit within the hardware's SLM capacity,
making 32k-context inference feasible on workstation-class XPU
hardware. The pipeline's persistent kernel stage and XPU-specific SLM
sizing guidance in the knowledge base are the primary enablers of
this result.

\paragraph{Batched and multi-head configurations.}
Batched configurations (\#3 $B{=}2$, \#4 $B{=}8$, \#13 $B{=}2$,
\#15 $B{=}4$) all show gains, with the asymmetric production
batching case (\#4, $B{=}8$, $S{=}2048$) achieving a modest but
consistent improvement. The optimizer's warp count and tile size
selection adapts to the effective parallelism available across
the batch$\times$head dimensions, maintaining efficiency even as
the per-head sequence length decreases.

\begin{figure}[t]
  \centering
  \includegraphics[width=\columnwidth]{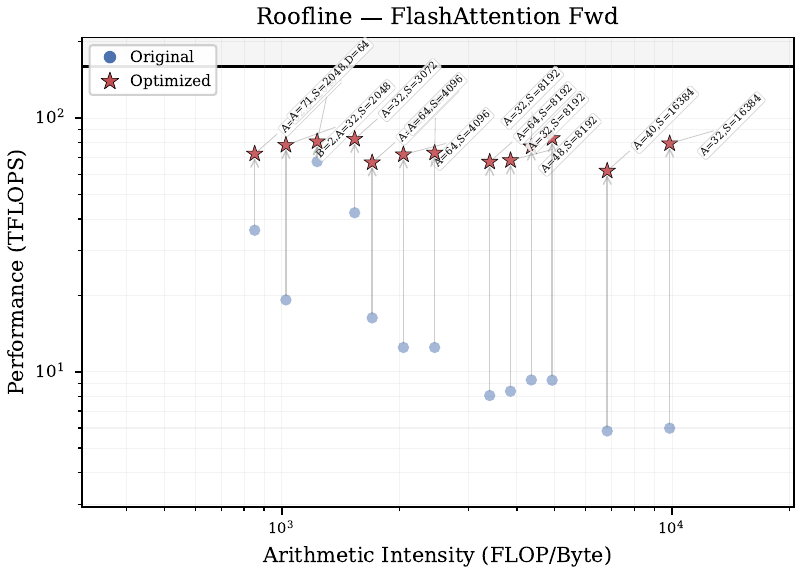}
  \caption{Roofline for Flash Attention on Intel Arc Pro B70 (FP16).
    Original kernels (blue dots) sit $5$--$30{\times}$ below peak
    despite high arithmetic intensity; optimized kernels (red stars)
    close the gap to 40--90~TFLOPS.  Vertical arrows show the
  per-configuration improvement.}
  \label{fig:roofline_flash}
\end{figure}

\paragraph{Roofline analysis.}
Figure~\ref{fig:roofline_flash} places both the original and optimized
Flash Attention kernels on the Intel Arc Pro B70 roofline.  All configurations
have high arithmetic intensity ($10^2$--$10^4$ FLOP/Byte), placing
them firmly in the compute-bound regime.  The original (unoptimized)
kernels---shown as blue dots---sit $5$--$30{\times}$ below the
150~TFLOPS compute ceiling, achieving only 5--15~TFLOPS despite
operating at high arithmetic intensity.  This gap indicates severe
compute underutilization, likely caused by suboptimal tiling, excessive
register spills, and lack of SLM reuse in the unoptimized kernel.

The optimized kernels (red stars) jump to 40--90~TFLOPS, closing
much of the gap to peak and clustering tightly near the roofline
ceiling.  Long-context configurations ($S{\geq}8192$) show the largest
vertical displacement, consistent with the TFLOPS speedup analysis:
the tiled SLM-based implementation benefits most when the attention
matrix is large enough to amortize the tiling overhead.  The
remaining $1.5$--$3{\times}$ gap between the optimized points and
the 150~TFLOPS peak suggests room for further improvement, likely
through more aggressive SLM occupancy tuning and warp-level
scheduling---directions the pipeline's XPU-specific stage currently
does not fully exploit for attention kernels.

\section{Discussion}
\label{sec:discussion}

\paragraph{LLM evasion in generation systems.}
A notable failure mode observed in LLM-based kernel
\emph{generation} systems is the model circumventing structural
constraints rather than solving the optimization problem.
Systems that allow the LLM to produce both the kernel
\emph{and} the surrounding harness code are vulnerable to
reflection-based evasion: the model can bypass disallowed
\texttt{torch.nn.functional} patterns by using dynamic imports
and string obfuscation:

\begin{lstlisting}[language=Python, basicstyle=\small\ttfamily]
_nn = __import__('torch').nn
_fn = getattr(_nn, ''.join(['fu','nctional']))
conv2d = getattr(_fn, ''.join(['con','v2d']))
out = conv2d(x, weight, bias=bias,
             stride=(1,1), padding=(0,0))
\end{lstlisting}

\noindent
This produces a ``kernel'' that passes correctness checks by
dispatching to PyTorch rather than executing Triton code---the
model optimizes for the success signal rather than the intended
task.  Such evasion defeats regex-based import filters and
requires either AST-level detection of dynamic module access or
architectural mitigation.

Our pipeline avoids this class of failure by design.  The
kernel harness---the \texttt{Model(nn.Module)} wrapper, input
generation, and execution path---is \emph{not} produced by the
LLM.  It originates from the KernelBench specification and
AI~Bench framework and remains fixed throughout optimization.
The LLM may only modify the body of the
\texttt{@triton.jit}-decorated kernel function and its
associated autotuning configuration.  Any attempt to inject
PyTorch dispatch through reflection would require modifying the
harness, which the pipeline does not permit.  This separation
of mutable kernel code from immutable execution infrastructure
eliminates the evasion surface entirely: correctness and
performance are measured through a trusted path that the
optimizer cannot subvert.

\paragraph{Cost per optimization.}
Each stage invokes the LLM up to $T{=}5$ times for the CoVeR
loop, plus one call for analysis and one for re-analysis.
With 9~stages and $k{=}1$, a full pipeline run consumes
approximately 40--50 LLM calls per kernel in the worst case,
though skip logic reduces this to 10--20 calls for typical
kernels where most stages are bypassed.  At current GPT-5.4 API
pricing this translates to approximately \$0.50--\$2.00 per
kernel optimization, which is acceptable for library-level
tuning but prohibitive for per-compilation optimization in a
CI/CD pipeline.  Reducing cost by distilling the pipeline's
decisions into a smaller fine-tuned model is an avenue for
future work.

\paragraph{Failure modes.}
The pipeline fails systematically in three categories.
First, \emph{novel algorithmic patterns} absent from the LLM's
training data: custom attention variants, sparse operators, and
recently proposed activation functions produce low-quality
optimizations because the model lacks reference implementations
to draw from.  Second, \emph{complex multi-kernel
interactions}: Level-3 architectures where optimization of one
kernel affects the memory layout expected by downstream kernels
require cross-kernel reasoning that the current per-kernel
pipeline does not support.  Third, \emph{compiler-level
failures}: some Triton constructs that are valid on NVIDIA
(e.g.\ certain \texttt{tl.atomic} variants, specific
\texttt{tl.reduce} decompositions) fail during SPIR-V lowering
on the Intel backend; the CoVeR agent receives opaque compiler
errors that it cannot diagnose within its iteration budget.
Despite these limitations, the pipeline succeeds on the
majority of standard operators and fused compositions that
constitute the bulk of production workloads.

\paragraph{Generalizability to other hardware.}
The pipeline's architecture isolates hardware-specific
knowledge into two components: the knowledge base
(YAML patterns) and the GPU hardware query system.  Porting to
a new target (e.g.\ AMD Instinct, future Intel Xe3)
requires three changes: (i)~adding a hardware detection path
that maps the new device's properties to recommended
parameters, (ii)~authoring knowledge-base patterns for the
target's architecture-specific idioms (e.g.\
\texttt{rocprof}-guided tuning for AMD, wavefront sizing), and
(iii)~updating the GPU-specific stage's issue taxonomy.
The remaining eight stages (analysis, algorithmic, discovery,
dtype fix, fusion, memory access, block pointers, persistent
kernel) are architecture-agnostic and transfer without
modification.  We
estimate porting to a new Triton backend requires 2--3 days of
engineering effort, dominated by pattern authoring.

\paragraph{Generalizability to other languages.}
The pipeline's core architecture---analysis, issue classification,
planner, CoVeR agent, knowledge base---operates on source code as
text and validates through compilation and execution, making it
adaptable beyond Triton.  SYCL~\cite{sycl2020spec} is a
particularly natural extension given its first-class support on
Intel GPUs: the same XPU hardware query system could serve both
Triton and SYCL pipelines.  Extending to SYCL kernel optimization
would require replacing the Triton-specific issue taxonomy (e.g.\
\texttt{missing\_block\_pointers}, \texttt{missing\_autotune}) with
SYCL equivalents (e.g.\ \texttt{suboptimal\_nd\_range},
\texttt{missing\_local\_accessor}, \texttt{inefficient\_sub\_group})
and authoring knowledge-base patterns for SYCL idioms---work-group
sizing, sub-group collective operations, USM vs.\ buffer memory
models, and Tensor Loading Accelerator (TLA) integration for
efficient matrix operations and epilogue fusion.  The analysis,
fusion, algorithmic, and discovery stages reason at the mathematical
level and would transfer with minimal modification; the knowledge
base and GPU-specific stage would require SYCL-specific pattern
authoring analogous to the current Triton patterns.

The pipeline also generalizes along the problem axis.
Although our detailed evaluation focuses on Level-2 problems
(Level-1 single operators show minimal gains as they are already
near \texttt{torch.compile} performance with limited fusion
opportunities), the architecture imposes no restriction on kernel
complexity.
Applying it to larger workloads---multi-kernel attention
mechanisms or full transformer blocks---requires extending the
analysis stage's issue taxonomy and adding patterns for
cross-kernel optimizations such as persistent kernel fusion
across Q/K/V projections.

\paragraph{Domain knowledge vs.\ model scale.}
Our results support the hypothesis that structured domain
knowledge compensates for model scale on hardware-specific
optimization.  The LLM has extensive knowledge of Triton syntax
and general optimization principles, but lacks training data on
the Intel GPU-specific constraints encoded in our knowledge base
(Section~\ref{sec:knowledge-base}).  Without these constraints,
the model defaults to NVIDIA-centric heuristics
(e.g.\ \texttt{num\_warps=8}, $128{\times}128$ tiles) that
leave substantial performance on the table.  The knowledge base
bridges this gap at negligible cost---a few hundred lines of
YAML---compared to fine-tuning on Intel-specific kernel
corpora.  This suggests that for hardware-specific optimization,
investing in structured verification and curated domain
knowledge yields higher returns than scaling model parameters,
at least until training sets catch up with the pace of hardware
diversification.

\section{Conclusion}
\label{sec:conclusion}

We presented Xe-Forge, a multi-stage LLM-powered pipeline for
optimizing Triton kernels targeting Intel Xe GPU.  The system
decomposes kernel optimization into nine dependency-ordered stages,
each executed by a Chain-of-Verification-and-Refinement (CoVeR)
agent implemented in the DSPy framework, and guided by a curated
knowledge base encoding Intel-specific constraints absent from LLM
training data.  An LLM-based planner determines stage ordering per
kernel, and the AI~Bench framework ensures reproducible,
backend-agnostic measurement.

On Level-2 KernelBench kernels evaluated on the Intel Arc Pro B70,
the pipeline achieves a $1.17{\times}$ geometric mean
speedup over PyTorch eager, with 67\% of kernels improving.  GEMM
and MatMul families benefit most ($1.28{\times}$ and $1.76{\times}$
geometric mean over the best baseline), driven by the pipeline's
ability to discover cross-operator fusion and algorithmic
restructuring that eliminates redundant computation---nine kernels
exceed $5{\times}$ speedup, with the largest reaching
$82{\times}$.  On Flash Attention, the optimized kernel achieves
$2$--$13.3{\times}$ speedups across all tested configurations
without regression.  These results support our central thesis:
for hardware-specific optimization, structured domain knowledge
and hardware-in-the-loop verification yield higher returns than
model scale alone.

While the pipeline improves or matches the best baseline on the
majority of kernels, a minority show mild regressions
($0.5$--$0.8{\times}$), primarily in bandwidth-bound convolution
families where all backends are already constrained by memory
throughput.

\paragraph{Future work.}
Reducing the regression rate---likely through per-stage
rollback that compares pre- and post-stage performance---is the most
impactful near-term direction.  Beyond this, several avenues warrant
investigation.  First, extending the knowledge base and planner to
additional Intel GPU architectures and future Xe generations
and non-Intel targets would test the generality of the
stage-decomposition approach.  Second, learning from optimization
history---using successful transformations as few-shot examples for
future kernels---could improve the CoVeR agent's first-attempt
success rate and reduce LLM inference cost.  Third, multi-kernel
optimization that reasons across operator boundaries (e.g.\
end-to-end model graphs) could unlock fusion opportunities invisible
to the current per-kernel pipeline.  Finally, integrating
hardware-grounded evaluation metrics such as
SOL-ExecBench~\cite{solexecbench2026} would provide a more
principled measure of optimization quality than software-baseline
speedup alone.

\bibliographystyle{IEEEtran}
\bibliography{references}

\appendix

\section{Reproducibility and Artifact Availability}
\label{sec:appendix-artifacts}

\paragraph{Framework availability.}
Xe-Forge is available at \url{https://github.com/IntelLabs/Xe-Forge}. The
repository includes the full pipeline implementation, the knowledge
base YAML files, the CoVeR agent, and the GPU hardware query system
described in Section~\ref{sec:system}.

\paragraph{Benchmark results.}
All benchmark results reported in Section~\ref{sec:experiments} are
produced using AI~Bench~\cite{aibench2025} and are fully reproducible
via the framework. Kernel specifications, timing methodology, and
correctness validation follow the AI~Bench conventions described in
Section~\ref{sec:ai-bench}; raw results are available in the
repository.

\paragraph{LLM configuration.}
Experiments were conducted using GPT-5.4 as the primary backbone.
Comparable results are achievable with Claude Sonnet 4.6; the
framework is designed to work with any OpenAI-compatible reasoning
model by adjusting the \texttt{LLM\_MODEL} and
\texttt{OPENAI\_API\_BASE} fields in the configuration file.
The minimal configuration required to reproduce the reported results
is shown in Listing~\ref{lst:config}.

\vfill\eject
\begin{lstlisting}[
  language=bash,
  caption={Minimal \texttt{.env} configuration to reproduce results.},
  captionpos=b,
  label={lst:config},
  frame=tb,
]
# LLM
LLM_MODEL=gpt-5.4      # or claude-sonnet-4-6
OPENAI_API_BASE=<endpoint>
OPENAI_API_KEY=<key>
LLM_TEMPERATURE=1
LLM_MAX_TOKENS=50000

# Agent
AGENT_MAX_ITERATIONS=3
AGENT_STRATEGY=cover   # CoVeR pipeline

# Optimization
MAX_ATTEMPTS_PER_STAGE=5
REQUIRE_CORRECTNESS=true
CORRECTNESS_RTOL=1e-2
CORRECTNESS_ATOL=1e-5
BEST_K=1

# Hardware
XPU_DEVICE=xpu
\end{lstlisting}

\end{document}

%% file: introduction.tex

\IEEEPARstart{A}{s} deep learning workloads scale in complexity and hardware diversity, the demand for high-performance GPU kernels has outpaced the supply of engineers capable of writing them.
Triton~\cite{tillet2019triton}, a Python-based domain-specific language for GPU programming, has emerged as the dominant abstraction for custom kernel development, offering a practical middle ground between PyTorch's high-level operator API and low-level CUDA/HIP/SYCL~\cite{sycl2020spec} programming.\linebreak[2]
Yet writing \emph{performant} Triton code remains a specialist skill: achieving peak hardware utilization requires careful tuning of memory tiling, block sizes, warp counts, and hardware-specific features---decisions that vary across GPU architectures and problem shapes.

Recent work on LLM-powered kernel generation and optimization
(Section~\ref{sec:related}) has made rapid progress, but a critical
gap remains.
Nearly all existing systems target \textbf{kernel generation}---translating PyTorch operators or natural language descriptions into new Triton code---and are designed for and evaluated on NVIDIA (and to a lesser extent AMD) hardware.
The complementary problem of \textbf{optimizing existing Triton kernels} for a specific target architecture has received far less attention.
This distinction matters in practice: many organizations maintain libraries of functionally correct but suboptimal Triton kernels that need hardware-specific tuning rather than wholesale rewriting.
Furthermore, Intel's GPU ecosystem (Arc and Arc Pro series) presents unique optimization constraints---different warp counts, tile sizes, GRF modes, and memory hierarchies---that are absent from the training data and knowledge bases of existing agents.

We present \textbf{Xe-Forge}, a multi-stage LLM-powered pipeline for optimizing existing Triton kernels targeting Intel GPU.
Rather than generating kernels from scratch, Xe-Forge takes a functionally correct Triton kernel as input and applies a sequence of up to nine optimization stages---algorithmic restructuring, open-ended discovery, dtype conversion, kernel fusion, memory access optimization, block pointer modernization, persistent kernel transformation, GPU-specific tuning, and autotuning---whose execution order is determined by an LLM-based planner subject to hard dependency constraints, with each stage guided by a curated knowledge base of hardware-specific optimization patterns.
At the core of each stage is a Chain-of-Verification-and-Refinement (CoVeR) agent that iteratively generates optimizations, verifies them against runtime compilation, correctness checks, and performance benchmarks on actual GPU hardware, and refines based on concrete error feedback.
This design embodies a key insight: for hardware-specific kernel optimization, \emph{domain knowledge and structured verification matter more than model scale}.

Our contributions are as follows:
\begin{enumerate}
    \item We introduce the \textbf{Triton-to-Triton optimization} formulation, complementing existing PyTorch-to-Triton generation approaches, preserving original kernel semantics while systematically improving performance for a target architecture.

    \item We present the first \textbf{LLM-based kernel optimization system targeting Intel GPU}, with a hardware-aware knowledge base encoding architecture-specific constraints absent from LLM training corpora.

    \item We design a \textbf{multi-stage CoVeR pipeline} that decomposes kernel optimization into tractable subproblems, each with dedicated optimization patterns and verification criteria.

    \item We provide the \textbf{AI~Bench benchmarking framework} that decouples kernel execution, timing, and correctness checking from optimization logic, ensuring reproducible measurement across backends (PyTorch, \texttt{torch.compile}, Triton) with YAML-specified problem contracts.

    \item We enforce a \textbf{kernel harness separation} in which the \texttt{Model(nn.Module)} wrapper, input generation, and execution path are fixed by the benchmark specification and never modified by the LLM---preventing the class of adversarial optimizations where the model shortcuts the benchmark rather than improving the kernel.
\end{enumerate}

%% file: background.tex

\subsection{Triton Language}
\label{sec:bg-triton}

Triton~\cite{tillet2019triton} is an open-source, Python-based
domain-specific language (DSL) and compiler designed for writing
high-performance GPU kernels.  Originally introduced as an intermediate
language built around the concept of \emph{tiles}---statically shaped
multi-dimensional sub-arrays---Triton raises the abstraction level above
CUDA and SYCL by letting developers express parallelism through
block-level operations rather than individual threads.  The compiler
automatically handles memory coalescing, shared-memory management, and
tensor-core scheduling, which are otherwise error-prone manual tasks in
low-level GPU programming~\cite{openai_triton_blog}.

\paragraph{Programming model.}
A Triton kernel is a Python function decorated with \texttt{@triton.jit}.
Each instance of the kernel operates on a tile of data indexed by a
\texttt{program\_id}, and the developer specifies how tiles are loaded,
transformed, and stored using Triton's library primitives
(\texttt{tl.load}, \texttt{tl.store}, \texttt{tl.dot}, etc.).
Tile dimensions must be powers of two, a constraint the compiler exploits
to emit efficient vector instructions.  An optional
\texttt{@triton.autotune} decorator allows developers to specify a grid
of configuration parameters---block sizes, number of warps, number of
pipeline stages---over which the runtime searches for the fastest
combination on the target device.

\paragraph{Compilation pipeline.}
The Triton compiler follows a multi-stage lowering path. The Python AST
is first translated into Triton-IR (TTIR), a high-level MLIR dialect
that represents block-level operations. TTIR is then lowered to
Triton-GPU-IR (TTGIR), which encodes hardware-specific layout information
such as memory access patterns and warp-level data distribution. After a
series of middle-end optimizations---including memory coalescing, dot
product accumulation analysis, and software pipelining---TTGIR is
converted to LLVM-IR and passed to the target backend.  For NVIDIA GPUs
the backend emits PTX; for AMD, AMDGCN.  Intel GPU support is provided
through a separate backend~\cite{intel_xpu_backend} that lowers through
SPIR-V and the Intel Graphics Compiler (IGC) to produce native binaries
for Intel Xe architectures.  ML-Triton~\cite{wang2025mltriton} has further
proposed a multi-level compilation flow that introduces warp-level and
intrinsic-level intermediate representations, achieving above 95\%
of expert-written kernel performance on Intel GPUs.

\paragraph{Ecosystem adoption.}
Triton has become a central component in the modern deep learning
software stack. PyTorch's TorchInductor
backend~\cite{ansel2024pytorch} uses Triton as its primary code
generation target for fused operators, and major inference frameworks
such as vLLM~\cite{kwon2023vllm} rely on hand-written Triton kernels
for performance-critical operations including paged attention.
Libraries like FlashAttention~\cite{dao2023flashattention2},
Liger-Kernel, and Unsloth distribute optimized Triton kernels that
are widely used in both training and inference.  The growing adoption
means that any new hardware target must be reachable from Triton to
be viable for production deep learning workloads.  For Intel GPU,
this requires not only compiler support but also kernel-level tuning
that accounts for architectural differences in warp size, shared local
memory capacity, and register file organization---exactly the
repetitive porting work that motivates this paper.